\documentclass[fleqn,usenatbib,useAMS]{mnras}
\usepackage{amsmath}	
\usepackage{amssymb}	
\usepackage{multicol}
\usepackage{bm}
\usepackage{float}
\usepackage{xcolor}
\usepackage[T1]{fontenc}
\usepackage{ae,aecompl}
\usepackage{newtxtext,newtxmath}
\usepackage{graphicx,epsfig}

\title[Analysis of modern astrometric catalogues in the {\it Gaia} era] {Analysis of modern astrometric catalogues in the {\it Gaia} era}
\author[V.~S. Akhmetov, P.~N. Fedorov, V.S. Tsvetkova and E.Yu. Bannikova] {
V.~S. Akhmetov$^{1}$\thanks{E-mail: akhmetovvs@gmail.com (VSA)},  P.~N. Fedorov$^{1}$, V.S. Tsvetkova$^{1,2}$ and E.Yu. Bannikova$^{2,1}$
\\
$^{1}$Institute of Astronomy, V.~N.~Karazin Kharkiv National University, Sumska St. 35, Kharkiv UA-61022, Ukraine\\
$^{2}$Institute of Radio Astronomy, National Academy of Sciences of Ukraine, Mystetstv 4, Kharkov, UA-61002, Ukraine}

\date{Accepted 2021. Received 2021; in original form 2021}

\pubyear{2021}

\begin{document}
\label{firstpage}
\pagerange{\pageref{firstpage}--\pageref{lastpage}}
\maketitle

\begin{abstract}
We investigate of the systems of proper motions of stars in the ground-based catalogues HSOY, UCAC5, GPS1 and PMA derived by combining with the {\it Gaia} DR1 space data. Assuming the systematic differences of stellar proper motions of two catalogues to be caused by the mutual solid-body rotation and glide of the coordinate systems produced by the data of the catalogues under comparison, we analyse the components of the mutual rotation vector and displacement of the origins of these systems. The equatorial components of the vector of mutual rotation velocity of the compared coordinate systems, as well as velocities of the mutual displacement of their origins, varying within the range from 0.2 to 2.9 mas yr$^{-1}$, were derived from a comparison of proper motions of the sources that are common for {\it Gaia} EDR3 and the TGAS, UCAC5, HSOY, GPS1 and  PMA catalogues, respectively.
The systematic errors of proper motions of stars in the HSOY, GPS1, PMA and {\it Gaia}~EDR3 catalogues in the range of faint stellar magnitudes were estimated by analysing the formal proper motions of extragalactic objects contained in these catalogues. The coordinate system realised by the {\it Gaia}~EDR3 data at the level of $<$ 0.1 mas yr$^{-1}$ is shown to have no rotation and glide relative to the LQAC-5, ALLWISEAGN, Milliquas extragalactis sources within the range from 15 to 21 stellar G magnitude. Among the ground-based catalogues, the system of proper motions of the PMA stars, which is independent of the  {\it Gaia}~EDR3 data, is the closest to the  {\it Gaia}~EDR3 system of proper motions in G magnitude range from 15 to 21.
\end{abstract}

\begin{keywords}
catalogues -- astrometry -- proper motions -- reference systems
\end{keywords}

\section{Introduction}

The first release of the {\it Gaia} astrometric Satellite data ({\it Gaia}~DR1), \citep {b1,b2} is known to contain 3 parameters: celestial coordinates and stellar magnitudes in the G band for about 1.1 billion sources, based on observations of the first 14 months of its operational phase. The TGAS catalogue \citep {m1} is the first large five-parameter astrometric catalogue of approximately 2 million objects of up to ~11.5 magnitude that contains positions and parallaxes taken from the {\it Gaia} data, as well as proper motions of stars derived from a combination of positions from the Hipparcos \citep{k1},\citep{l1} and Tycho-2 \citep{h1} catalogues. During one year, this has stimulated the creation of four new astrometric catalogues, which contain the proper motions of stars: HSOY \citep{a3}, UCAC5 \citep{z1}, GPS1 \citep{t1} and PMA \citep{a1}. When creating these catalogues, data from the {\it Gaia} DR1 catalogue were used to one degree or another.

When creating the UCAC-5 catalogue, the TGAS catalogue was used as the reference one to reduce the US Naval Observatory CCD Astrograph catalogue (UCAC) observational data into the {\it Gaia} DR1 reference frame. This made it possible to obtain proper motions for more than 107 million stars with a typical accuracy of ±1 to ±2 mas yr$^{-1}$ in the range from 11 up to 15 R magnitude, and of approximately ±5 mas yr$^{-1}$ for 16 magnitude \citep{z1}.

The HSOY catalogue was derived by combining positions from PPMXL \citep{r1} and {\it Gaia} DR1 with the use of the weighted least squares method, which has been applied in deriving the PPMXL catalogue itself. The catalogue contains 583 million stars with the position accuracy close to that of {\it Gaia} DR1, and proper motions accurate to approximately ±1 to ±5 mas yr$^{-1}$, depending on stellar magnitude and coordinates of objects in the sky.

In fact, a classic method was applied to create these two catalogues, when the reference frame specified in the bright part of the magnitude range is used through the entire range, including its faint part. The lack of knowledge about the magnitude equation in the entire range of magnitudes is a significant problem, which usually leads to systematic errors in the system of proper motions.

A combination of astrometric data from {\it Gaia} DR1, Pan-STARRS DR1 \citep{c1}, Sloan Digital Sky Survey \citep{g1} and 2MASS \citep{s2} allowed to determine proper motions for 350 million sources in the GPS1 catalogue that covers three quarters of the sky. The GPS1 catalogue is presented in the form of two datasets - GPS1a and GPS1b, which differ in stellar proper motion systems due to the use of various combinations of astrometric data from {\it Gaia} DR1, Pan-STARRS DR1, SDSS and 2MASS. The authors claim that the GPS1 catalogue has a characteristic systematic error of less than 0.3 mas yr$^{-1}$ and a typical random uncertainty of 1.5 – 2.0 mas yr$^{-1}$ \citep{t1}.

The PMA catalogue was created by combining the 2MASS \citep{s2} and {\it Gaia} DR1 \citep{b1} positions. The catalogue contains positions of more than 420 million objects from {\it Gaia} DR1, and the absolute (independent of {\it Gaia} DR1) proper motions, which cover the whole celestial sphere without gaps for the magnitude range from 8 to 21 magnitude. The absolute calibration procedure (establishing the zero-point of proper motions) was performed with the use of about 1.6 million extragalactic sources extracted from the {\it Gaia} DR1 data according to a special technique \citep{f1}. The mean formal error of the absolute calibration is less than 0.35 mas yr$^{-1}$. The root mean square error of proper motions depends on stellar magnitude and ranges within $\pm(2 - 5)$ mas yr$^{-1}$ for the stars with 10 to 17 G magnitude and $\pm(5 - 10)$ mas yr$^{-1}$ for fainter stars \citep{a1}.

Note that the proper motions of sources in the PMA and GPS1 catalogues were derived not by the classical method that was proposed when creating the XPM catalogue \citep{f3} and \citep{f4}. Zero-points of proper motions of stars in these catalogues were established using the positions of extragalactic sources from {\it Gaia} DR1. In contrast to the HSOY and UCAC-5 catalogues, the reference system in the PMA and GPS1 catalogues, specified by the time-independent positions of galaxies, extends into the bright part of the magnitude range.

The second release of the {\it Gaia} data {\it Gaia}~DR2, \citep{b3} provides complete astrometric data (positions, parallaxes and proper motions) for over 1.3 billion sources, which contain more than 550 000 quasars. These sources were used in the astrometric solution for {\it Gaia}~DR2 with the aim to avoid rotation and alignment of the axes with the prototype version of the forthcoming third implementation of ICRF \citep{m2}. Thus, the positions of weak quasars constitute the main realisation of {\it Gaia}-CRF2, while the positions and proper motions of 1.3 billion stars are nominally in the same {\it Gaia}-CRF2 frame and provide a secondary realisation - the {\it Gaia} DR2 stellar reference frame. This reference frame covers the range of G magnitudes from 6 to 21 magnitude with similar accuracy, but deteriorates with increasing time span from the reference epoch J2015.5.

The third data release of the {\it Gaia} is splite into two installments: the early release called Gaia Early Data Release ({\it Gaia}~EDR3 with new astrometric solution for 1,811,709,771 objects that based on 34 months observation \citep{l3} and 1 614 173 {\it Gaia}-CRF3 source and also astrophysical parameters from {\it Gaia}~DR2), and the full Gaia Data Release 3 ({\it Gaia}~DR3) that is planed for the first half of 2022 \citep{b4}.

The main goal of the present work is to analyse the systems of proper motions of the TGAS, UCAC5, HSOY, GPS1 and PMA catalogues and to compare them with the system of proper motions specified by the  {\it Gaia}~EDR3 catalogue. In Section ~\ref{Section2} we determine the components of the angular velocity of rotation of the coordinate systems, as well as the components of the velocity of displacement of the origins of coordinate systems, specified by the TGAS, UCAC5, HSOY, GPS1 and PMA catalogues relative to the {\it Gaia}~EDR3 stellar reference frame. In Section ~\ref{Section3} we determine the same parameters based on the analysis of formal proper motions of extragalactic sources contained in these catalogues.

\section{Comparison of {\it Gaia}~EDR3 with the data of other catalogues}
 \label{Section2}
To analyse the proper motions of stars presented in two catalogues, we use a model of their systematic differences, assuming that they are generated by mutual solid-body rotation and displacement of two coordinate systems implemented by the data of these catalogues. In our consideration, the equations that describe these two effects in systematic differences are based on transformation of the coordinate system corresponding to the {\it CAT} catalogue into the {\it Gaia}~EDR3 coordinate system using the origin offset and rotation of the axes by orientation angles. In the approximation of infinitely small orientation angles, which depend on time, the equations for the differences in the proper motions of stars in right ascension ($\Delta\mu_{\alpha}$) and declination ($\Delta\mu_{\delta}$) are usually represented in the form:

\begin{equation}\label{eq1}
\begin{array}{lcl}
(\mu_{\alpha}^{CAT}-\mu_{\alpha}^{GaiaEDR3})\cos{\delta}=\omega_{x}\sin{\delta}\cos{\alpha} +\\+ \omega_{y}\sin{\delta}\sin{\alpha}- \omega_{z} \cos{\delta } + \dot{g}_{x} \sin{\alpha}  -  \dot{g}_{y} \cos {\alpha}, 
\end{array}
\end{equation}
\begin{equation}\label{eq2}
\begin{array}{lc}
(\mu_{\delta}^{CAT}-\mu_{\delta}^{GaiaEDR3})=-\omega_{x}\sin{\alpha} + \omega_{y}\cos{\alpha}\\ + \dot{g}_{x}\cos{\alpha}\sin{\delta} + \dot{g}_{y}\sin{\alpha}\sin{\delta} - \dot{g}_{z}\cos{\delta},
\end{array}
\end{equation}
where $\omega_{x},\omega_{y}$ and $\omega_{z}$ are components of the angular velocity vector (AVV) expressed in mas yr$^{-1}$ in the Cartesian coordinate system specified by the CAT catalogue relative to that one specified by the {\it Gaia}~EDR3 catalogue. $\dot{g}_{x}$, $\dot{g}_{y}$, $\dot{g}_{z}$ are the components of the displacement velocity vector (DVV) of the origin of the CAT catalogue coordinate system relative to the origin of the {\it Gaia}~EDR3 system. If we neglect the components $\dot{g}_{x}$, $\dot{g}_{y}$, $\dot{g}_{z}$ in equations (1) and (2), we get the equations that have been used \citep{l2} when deriving the Hipparcos catalogue \citep{k1} and \citep{l1}. The effect of the $\dot{g}_{x}$, $\dot{g}_{y}$, $\dot{g}_{z}$ components on the systematic differences was revealed in \citep{m3}, where it is considered as a flow, or a continuous glide from one point to another, which are located diametrically opposite on the sphere.

To identify common stars in the catalogues and to obtain the differences of their proper motions, we performed cross-identification in a circular search window with a radius of 0.5 arcseconds for (TGAS –  {\it Gaia}~EDR3), (UCAC5 –  {\it Gaia}~EDR3), (HSOY–  {\it Gaia}~EDR3), (GPS1 – {\it Gaia}~EDR3) and (PMA –  {\it Gaia}~EDR3) using method that described in \citep{a2}.

\begin{figure*}
 \includegraphics[width = 88mm,]{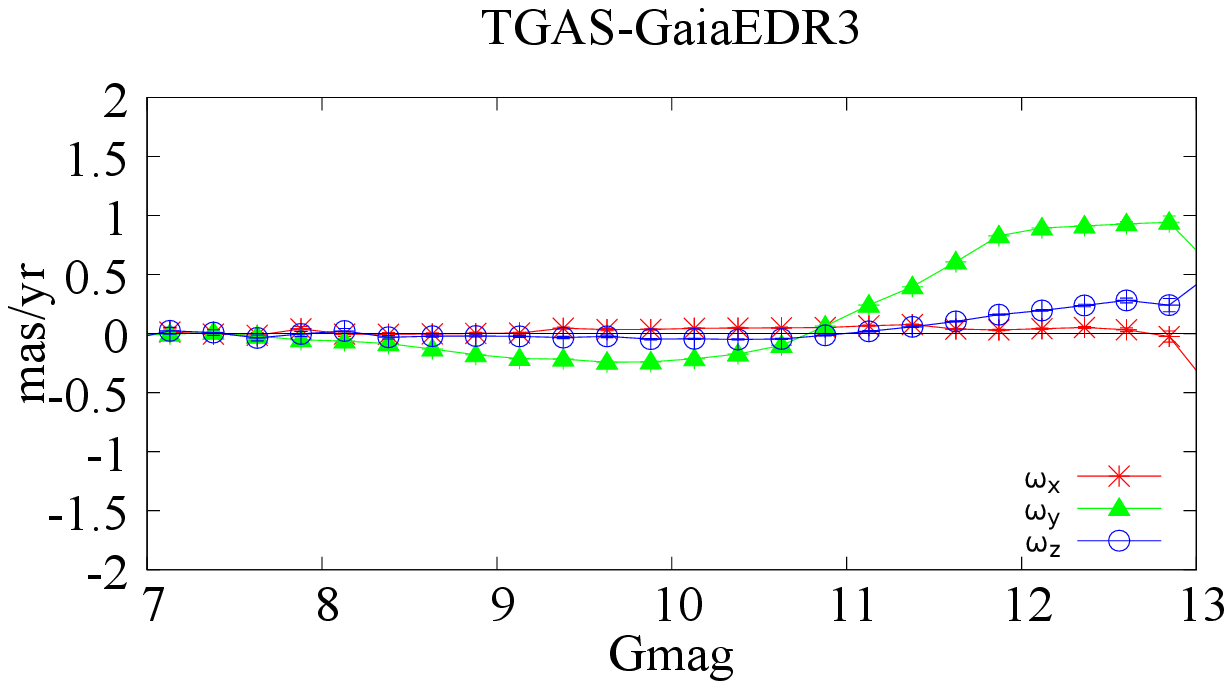}
 \includegraphics[width = 88mm,]{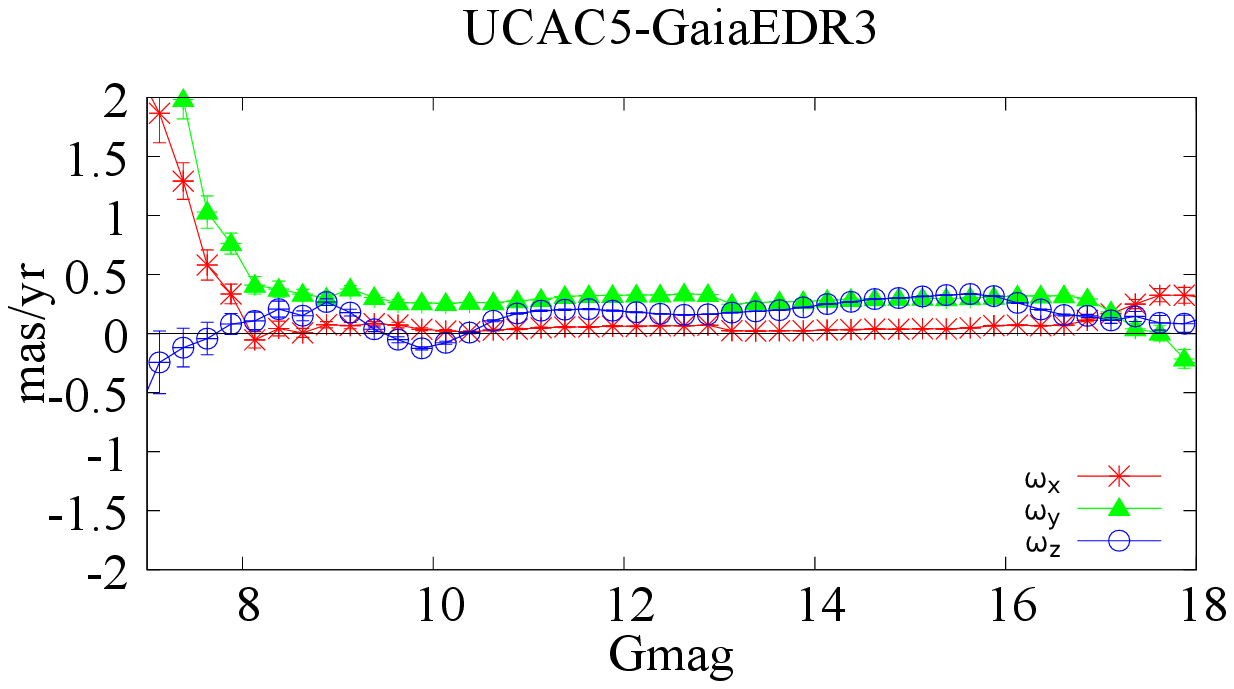}
 \includegraphics[width = 88mm,]{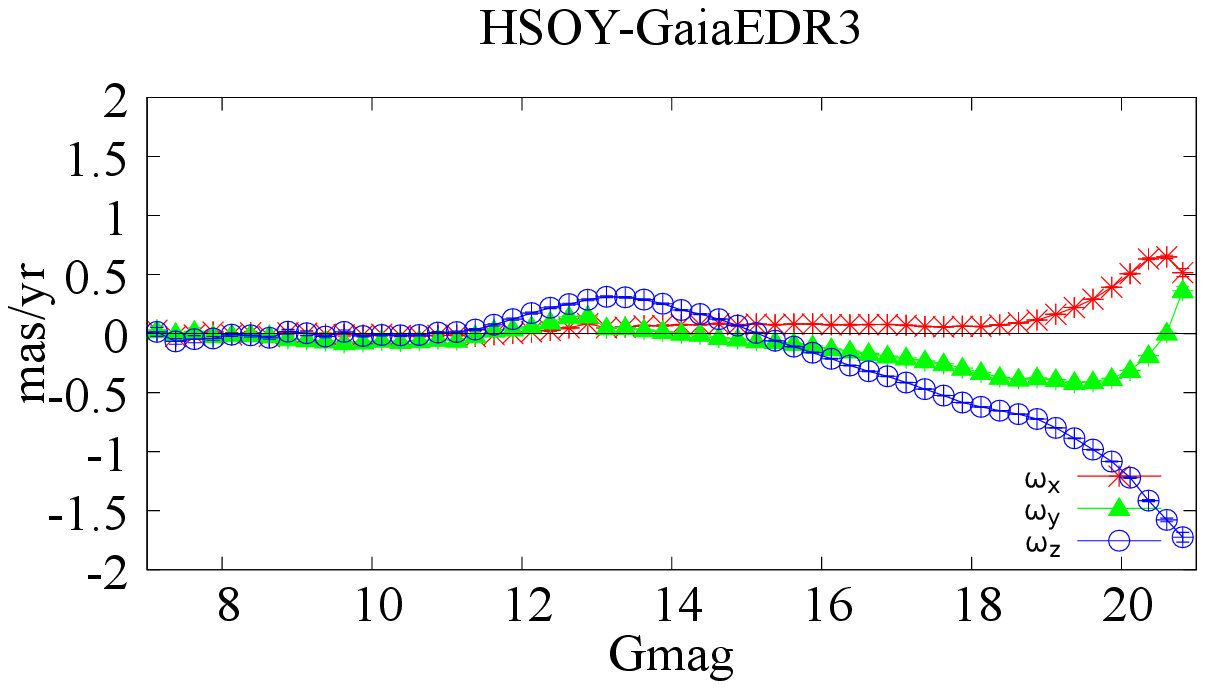}
 \includegraphics[width = 88mm,]{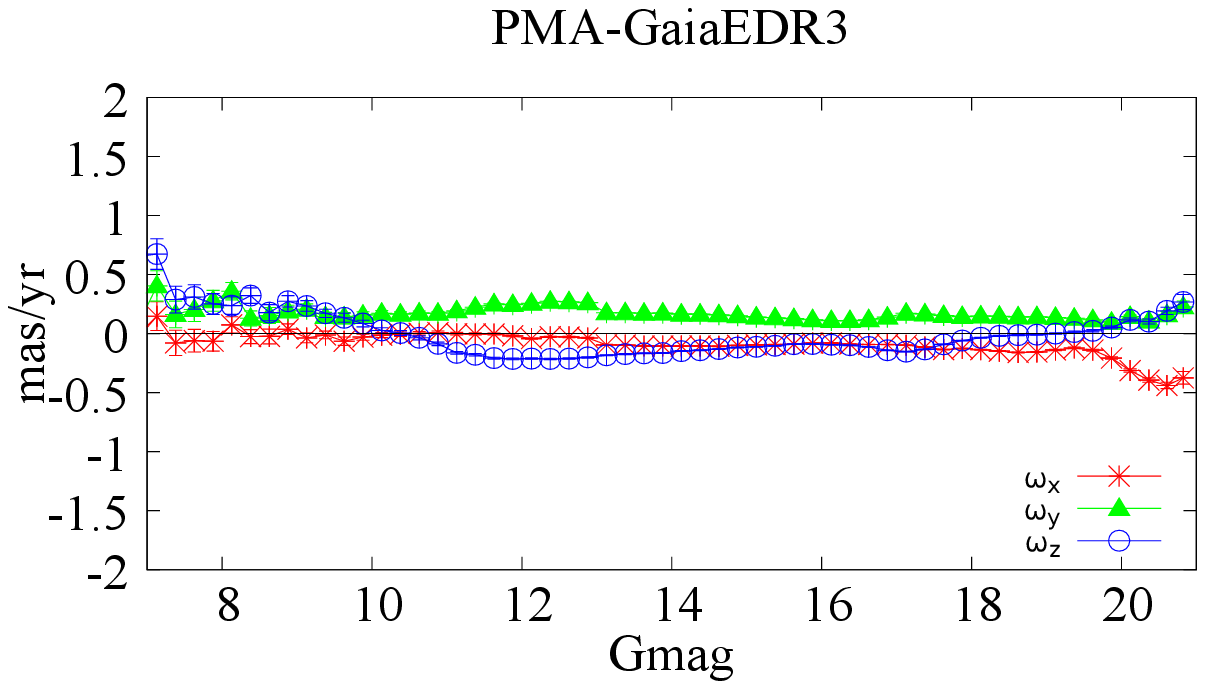}
 \includegraphics[width = 88mm,]{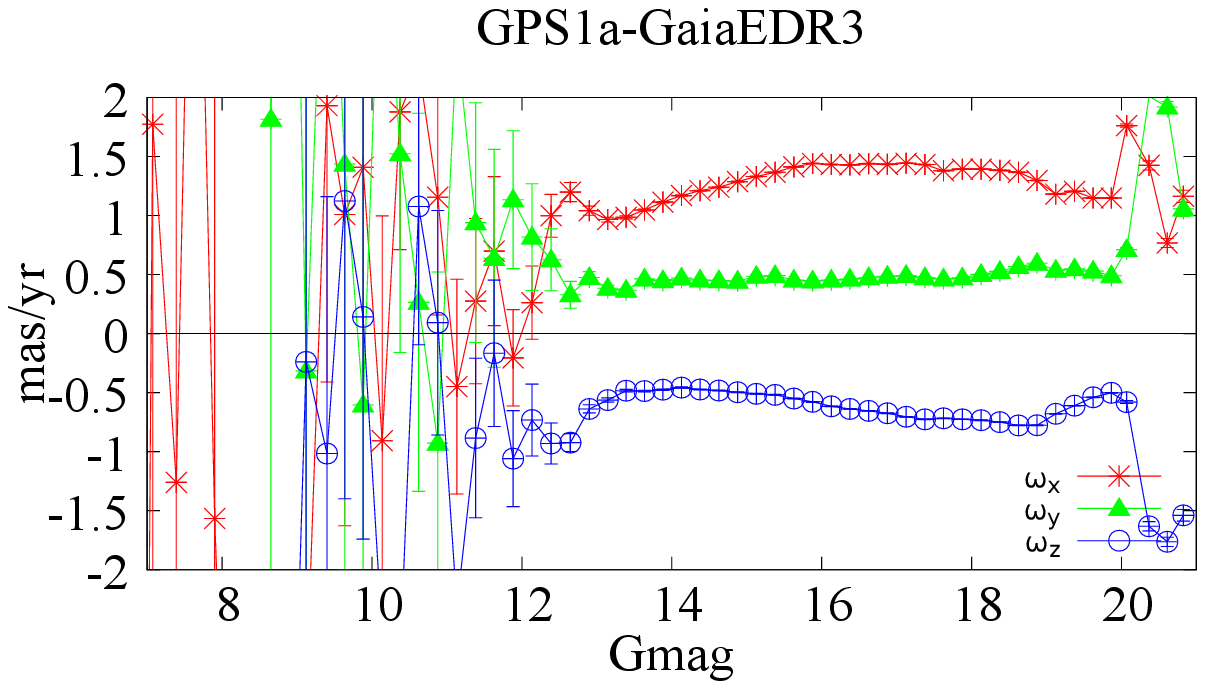}
 \includegraphics[width = 88mm,]{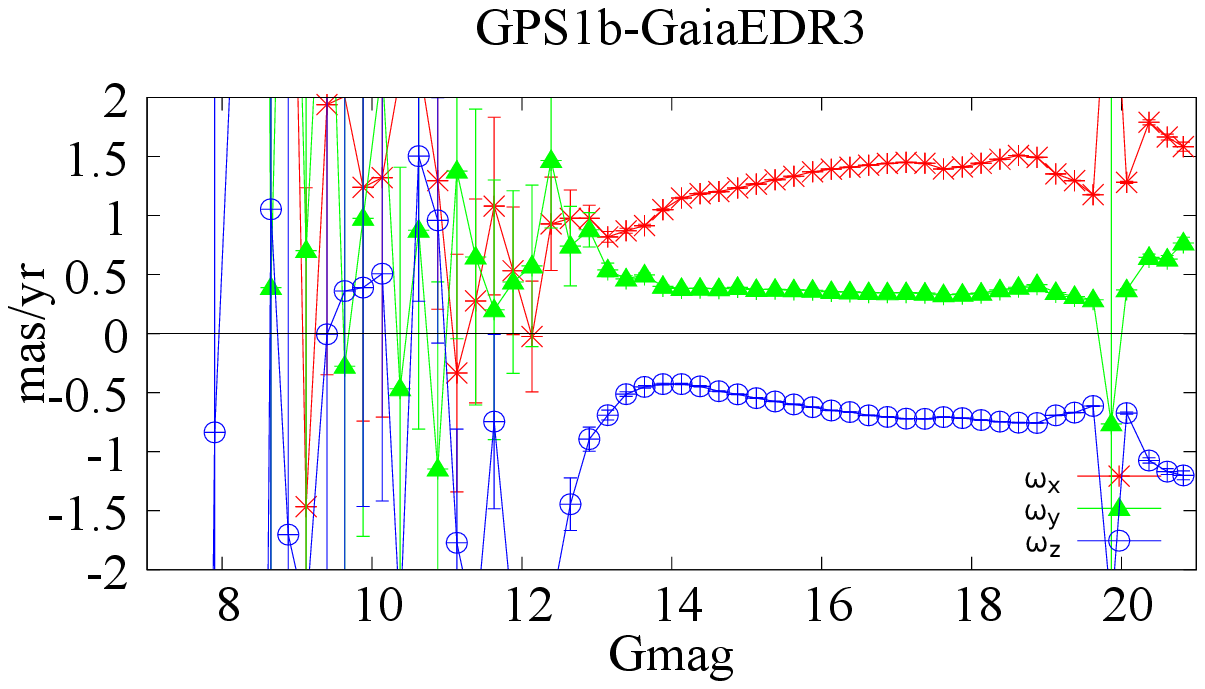}
\caption{Components $\omega_{x}$ (asterisk),  $\omega_{y}$(triangles)  and $\omega_{z}$ (open circles) of the mutual rotation vector between the TGAS, UCAC5, HSOY, PMA, GPS1a, GPS1b coordinate systems and the {\it Gaia}~EDR3 system as a function of G stellar magnitude from {\it Gaia}EDR3.}
\label{fig1}
\end{figure*}

\begin{figure*}
\includegraphics[width = 88mm,]{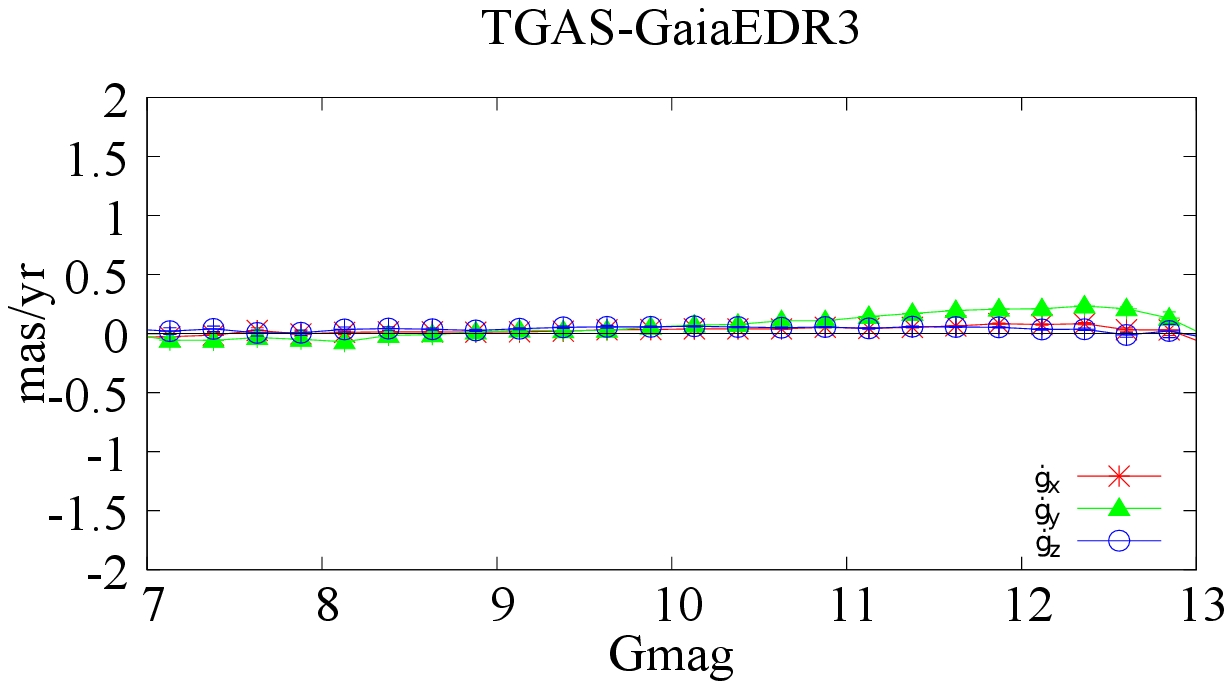}
\includegraphics[width = 88mm,]{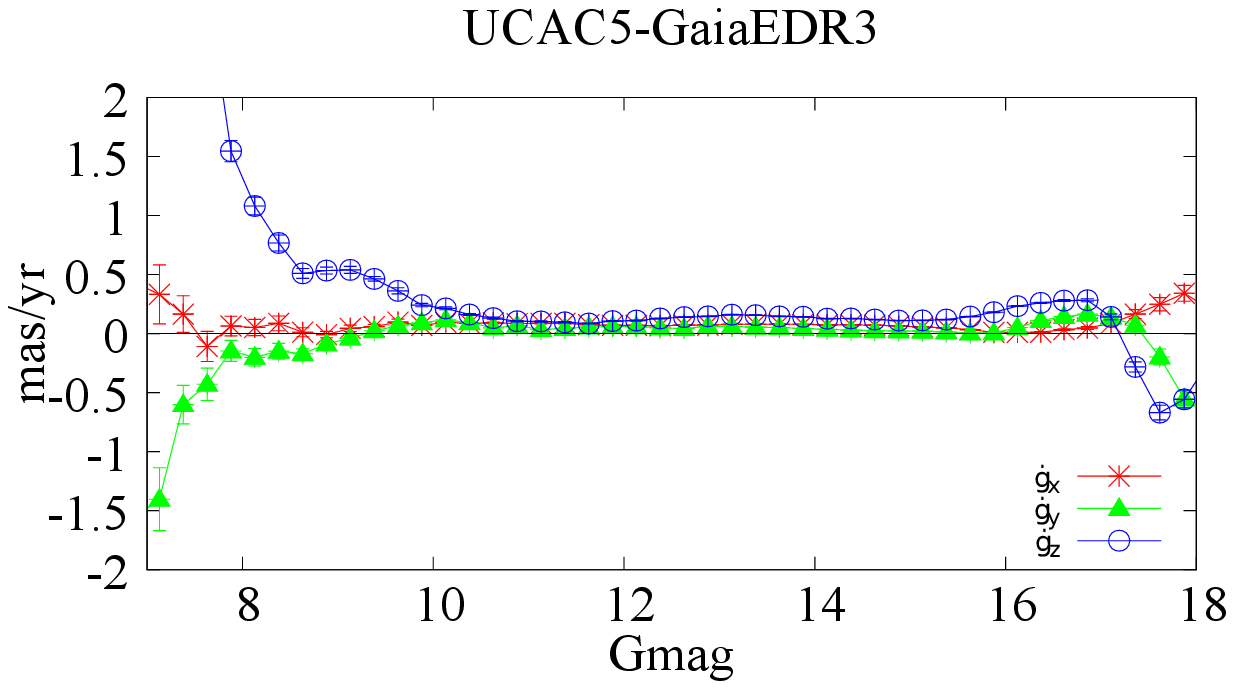}
\includegraphics[width = 88mm,]{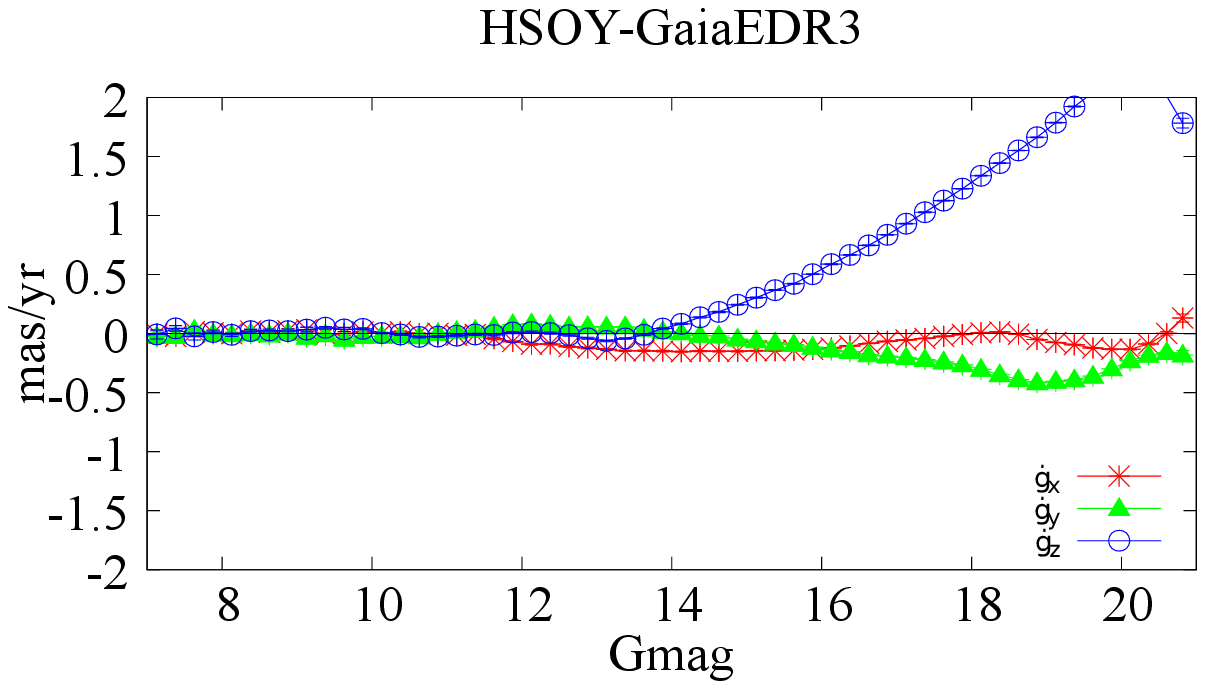}
\includegraphics[width = 88mm,]{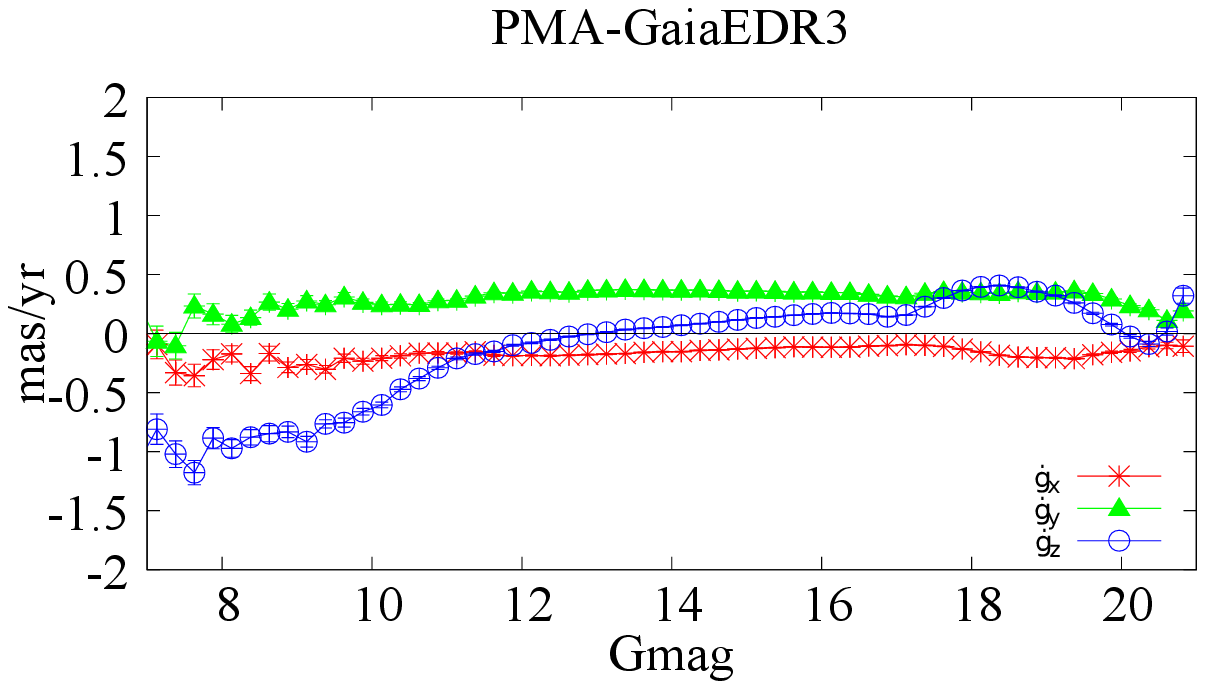}
\includegraphics[width = 88mm,]{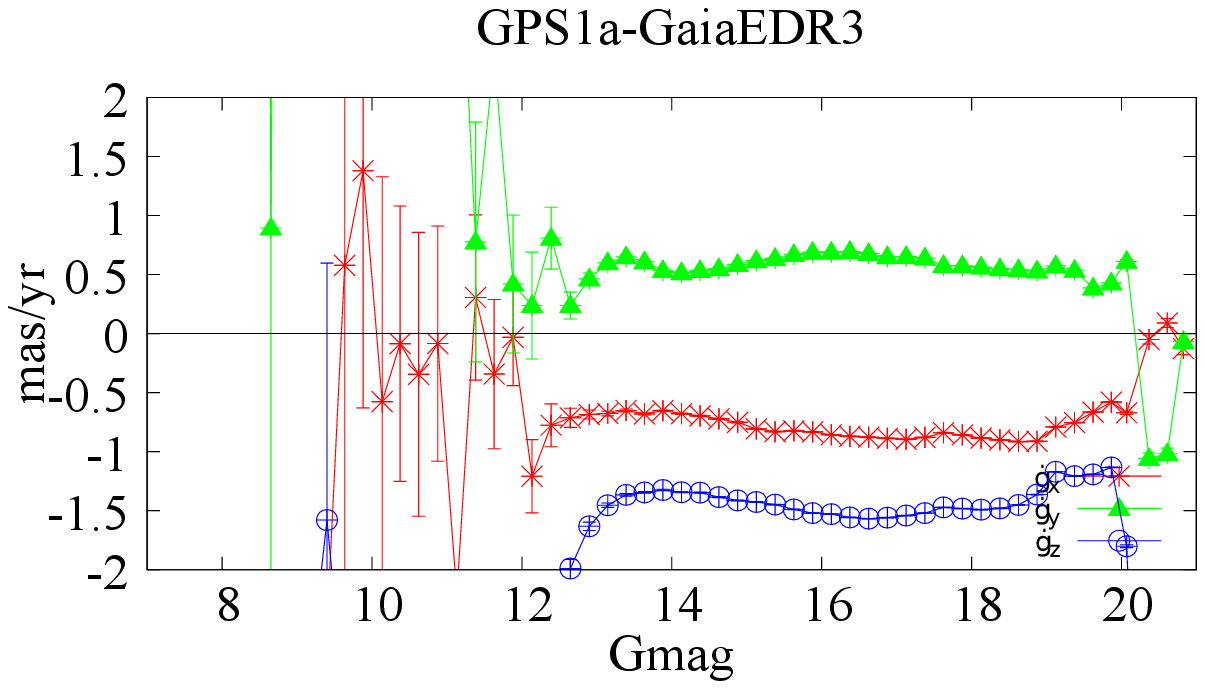}
\includegraphics[width = 88mm,]{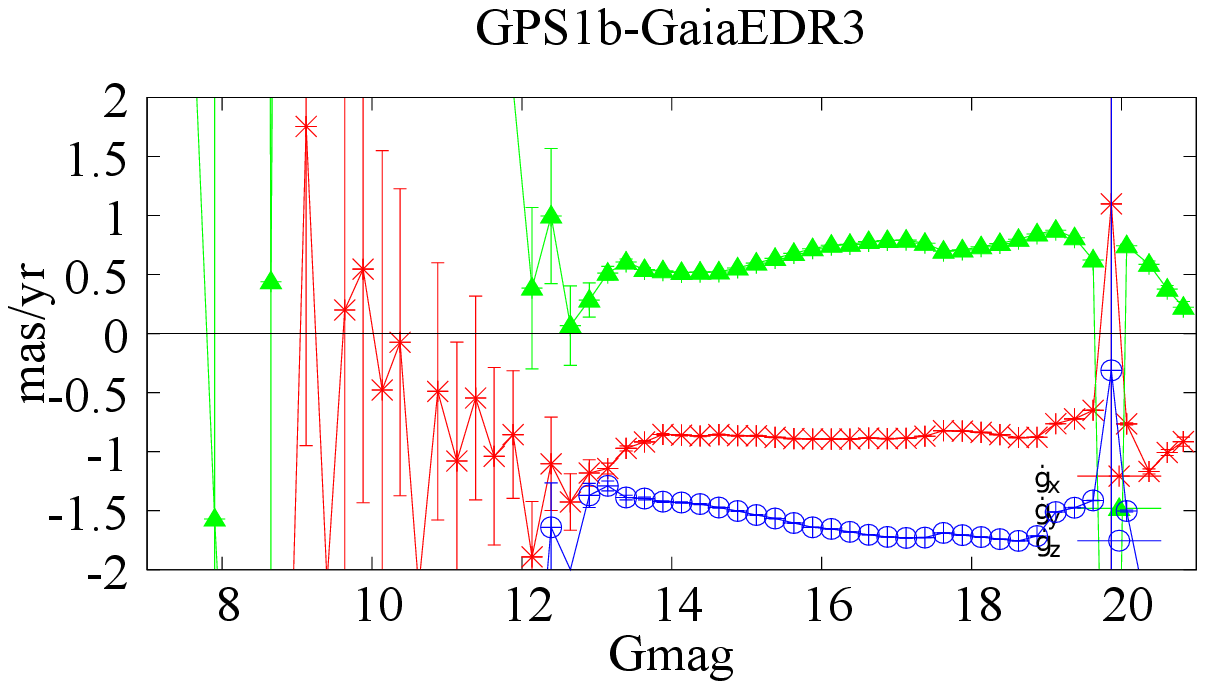}
\caption{Components $\dot{g}_{x}$ (asterisk), $\dot{g}_{y}$ (triangles)  and $\dot{g}_{z}$ (open circles) of the displacement velocity vector between the TGAS, UCAC5, HSOY, PMA, GPS1a, GPS1b coordinate systems and the {\it Gaia}~EDR3 system as a function of G stellar magnitude from {\it Gaia}EDR3.}
\label{fig2}
\end{figure*}

In Fig.~\ref{fig1} and Fig.~\ref{fig2} the AVV and DVV components are presented for the coordinate systems, which are specified by catalogues TGAS, UCAC5, HSOY, GPS1 and PMA relative to the system specified by the {\it Gaia}~EDR3 catalogue.
In addition, \citep{f2} have shown on the basis of the {\it Gaia}~DR1 data that proper motions of stars of the TGAS catalogue contain a systematic error. This error, which manifests itself most strongly in component $\omega_{y}$. At the same time, no such error was found in the proper motions of the Hipparcos stars from the TGAS catalogue, at least in the range from 6 to 9 stellar G magnitude. Obviously, the values of the detected errors are not negligible; however, for users who have the {\it Gaia}~DR2 and {\it Gaia}~EDR3 stellar reference frame at their disposal today, and the final release of the {\it Gaia} catalogue in the near future, the TGAS catalogue is unlikely to be of interest.

As we have already noted in the Introduction, the coordinate systems specified by the UCAC5 and HSOY catalogues, due to the principles of their construction, must be very close to that specified by the {\it Gaia}~EDR3 catalogue and must not have mutual rotation. This is indeed the case for the UCAC5 catalogue, for which the AVV and DVV components relative to the {\it Gaia}~EDR3 frame are practically zero at the accuracy level  $\pm(0.1 - 0.3)$ mas yr$^{-1}$ throughout the entire range of stellar magnitudes.

At the same time, in the bright part of the magnitude range, the coordinate system specified by the HSOY catalogue also has no rotation and glide relative to the {\it Gaia}~EDR3 coordinate system. This is quite expectable, as the reduction of the PPMXL data was carried out with the use of the reference TGAS catalogue that contains proper motions of stars up to 11.5 G magnitude. However, starting from 11 G magnitude, the AVV and DVV components begin to increase and reach several mas yr$^{-1}$ at 20 G magnitude. This indicates the presence of noticeable systematic errors in the system of proper motions of stars in the HSOY catalogue. The magnitude dependence of the AVV and DVV components indicates that the proper motions of HSOY stars are burdened by significant effects of the magnitude equation.

The catalogues built on the same principle as UCAC5 and HSOY, were necessary and useful to propagate into the weak part of the magnitude range and to condense a coordinate system, well defined in the bright part of the magnitude range. However, such catalogues, on one hand, do not allow obtaining any information on systematic errors of the reference catalogue, and on the other hand, do not ensure the absence of systematic errors outside the magnitude range of the reference catalogue.

The systems of proper motions of stars of the PMA and GPS1 catalogues, by the principle of their construction, are independent on the {\it Gaia}~EDR3 system. This feature make them useful for testing the compared reference systems and estimating the measure of their quasi-inertiality.

As seen from Fig.~\ref{fig1}, the AVV components between the systems implemented in the UCAC5 and {\it Gaia}~EDR3 catalogues do not exceed 0.3 mas yr$^{-1}$ up to 17 magnitude, and are in good agreement with the behavior of $\omega_{x}$, $\omega_{y}$, and $\omega_{z}$ obtained from a comparison of the {\it Gaia}~EDR3 and PMA catalogues. Note that components $\omega_{x}$, $\omega_{y}$ and $\omega_{z}$, which characterize the rotation between the {\it Gaia}~EDR3 and PMA systems, do not exceed 0.5 mas yr$^{-1}$ in the weaker part of the magnitude range as well, up to 20 magnitude. It can be said that the systems of proper motions of the UCAC5 and PMA catalogues in their common magnitude range are rather close at the level of 0.3 - 0.5 mas yr$^{-1}$. In fact, the same conclusion follows from consideration of the behaviour of the DVV components of origins of the coordinate systems implemented by these catalogues relative to the {\it Gaia}~EDR3 coordinate system (see Fig.~\ref{fig2}).

At the same time, the values and behaviour of the AVV and DVV components ( see Fig.~\ref{fig1} and Fig.~\ref{fig2} ) obtained by comparing the GPS1 and {\it Gaia}~EDR3 catalogues, noticeably differ from those obtained by comparing the PMA catalogue with {\it Gaia}~EDR3. Since the system of proper motions of stars from the PMA catalogue turned out to be close to that of the {\it Gaia}~EDR3 stars, we state that the system of proper motions of stars from the GPS1 catalogue has noticeable systematic differences from those of the {\it Gaia}~EDR3 and PMA catalogues. We do not assert here that the detected differences in the AVV and DVV components are a consequence of systematic errors in the proper motions of the GPS1a and GPS1b stars. The conclusion about the quality of the GPS1a and GPS1b system of proper motions of stars will be finally made in Section \label{Section3}.

\section{Analysis of the formal proper motions of extragalactic objects}
\label{Section3}

The rotation velocity and the velocity of displacement of the coordinate systems realised by the above noted catalogues can also be estimated by analysing the formal proper motions of extragalactic objects contained in the catalogues. Such formal proper motions of extragalactic sources are listed only in the {\it Gaia} DR3, PMA, HSOY and GPS1 catalogues. Since extragalactic sources are very distant objects, their observed proper motions, being very small, virtually do not change direction toward them from the barycenter of the solar system. The change in direction caused by the motion of the barycenter around the center of the Galaxy can be neglected in this case, since the time interval for observing these objects is also extremely short.

If the proper motions of objects of a catalogue were derived within the same procedure, without dividing into galactic and extragalactic sources, then the systems of proper motions of stars and extragalactic sources must demonstrate consistency. This means that the coordinate system set for an arbitrary epoch with the use of positions and proper motions of stars, must coincide with the coordinate system set by the directions to extragalactic sources. Thus, the formal proper motions of extragalactic objects in a catalogue are an excellent indicator of quasi-inertiality of the coordinate system, realized by positions and proper motions of stars.

Usually, the solution of the problem of assessing the degree of inertiality of the coordinate system implemented by the data of the catalogue under consideration is also carried out by finding the AVV and DVV components. Obviously, when using extragalactic sources, the AVV and DVV components determined from their formal proper motions should be expected to be close to zero. The nonzero values of the proper motions of extragalactic sources at the current level of accuracy are mainly due to random errors of their measurements, which, given a relatively uniform distribution over the celestial sphere, cannot cause a systematic bias in the estimates of AVV and DVV. In the case of nonzero values of AVV and DVV, the axes of the coordinate system implemented by the stellar data of a particular catalogue can be argued to have rotation, and the origin of coordinates has the displacement velocity. It is obvious that these effects are caused by the presence of systematic errors in the system of proper motions of the catalogue’s objects.

To solve this problem, it is necessary to identify the extragalactic sources contained in the catalogues under examination. For identification, we used catalogues of extragalactic sources AllWISEAGN, Milliquas, and LQAC5, which cover the entire celestial sphere.

The AllWISEAGN catalogue present an all-sky sample of $\approx$~1.4 million active galactic nuclei (AGNs) \citep{s1}. This catalogue was obtained from observations by a wide-angle infrared survey explorer, WISE \citep{w1}, which operates in the mid-IR wavelengths of 3.4, 4.6, 12 and 22 $\mu$m. The AllWISE AGN catalogue has relatively uniform sky coverage, with the exception of the Galaxy plane and the ecliptic poles. The sources are classified as AGNs according to the two-color criterion, and, as the authors claim, probability of contamination by stars is $4.0\times10^{-5}$ for a source. About half of the AllWISE AGN sources have optical counterparts that were detected by {\it Gaia} at least once during the first two years.

Million Quasars (Milliquas) is a catalogue of 607,208 Type I QSOs and AGNs that is compiled primarily from literature data of up to August 5, 2017, including the SDSS-DR14 release \citep{f5}. Also included are 1.35 million quasar candidates from photometric quasar catalogues NBCKDE, NBCKDE-v3, XDQSO, AllWISE and from all-sky radio/X-ray associated objects. Type II and Bl Lac objects are also included, resulting in the total number of 1,998,464 \citep{f5}. 

The final LQAC-5 catalogue contains 592,809 quasars. Among them, 398,697 objects were found in the {\it Gaia}~DR2 catalogue in a search window with the radius of 1 arcsecond. For the first time, parallaxes and proper motions of objects available in {\it Gaia}~EDR3 were added to the compilation. LQAC-5 is an almost complete catalogue of spectroscopically confirmed quasars (including a small fraction of 14,126 compact AGNs) that provides their best equatorial coordinates relative to ICRF2, and additional information about their physical properties, such as redshifts, multiple-band (photometric data) apparent and absolute magnitudes \citep{s4}. Also, cross-identification of the LQAC-5 catalogue with the AllWISE survey provides additional information for the mid-IR range.

Using cross-identification of the AllWISEAGN, Milliquas, LQAC5 catalogues sequentially with {\it Gaia}~EDR3, HSOY, GPS1 and PMA in a 1.0 arcsecond search window, and with the subsequent application of filters based on the {\it Gaia} astrometry (see Section 5.1, equation (13) in \citep{b3}), we obtained the lists of identified extragalactic objects. The filters selected sources with good observational records, formal parallax uncertainty < 1 mas, reliable significance level of parallax and proper motion from the region of the Galaxy, where the condition $|\sin b|$ > 0.1 is satisfied. The number of extragalactic sources from AllWISEAGN, Milliquas, and LQAC5 after sequential cross-identification with the {\it Gaia}~EDR3, HSOY, GPS1, and PMA catalogues and subsequent filtering is presented in Table ~\ref{table1}.
 
\begin{table}
\centering
\caption{Number of extragalaxies objects in the modern height density astrometric catalogues}
\label{table1}
\begin{tabular}{ | l | l | c | c |}
\hline
Catalogue & AllWISEAGN & Milliquas & LQAC5  \\ \hline
{\it Gaia}EDR3 & 671083 & 1278780 & 414252 \\ \hline
HSOY & 492956 & 811523 & 271512 \\ \hline
GPS1 & 315264 & 467689 & 178266 \\ \hline
PMA & 84615 & 83458 & 28602 \\ \hline
\end{tabular}
\end{table}

\begin{figure*}
 \includegraphics[width = 58mm,]{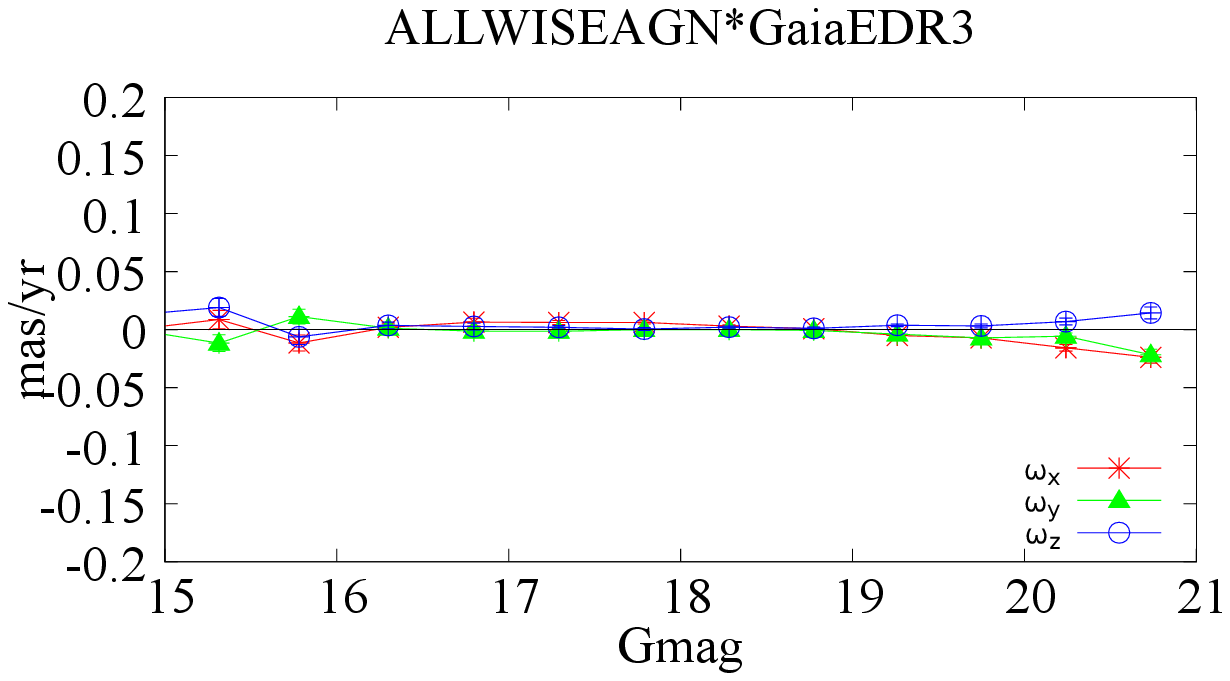}
 \includegraphics[width = 58mm,]{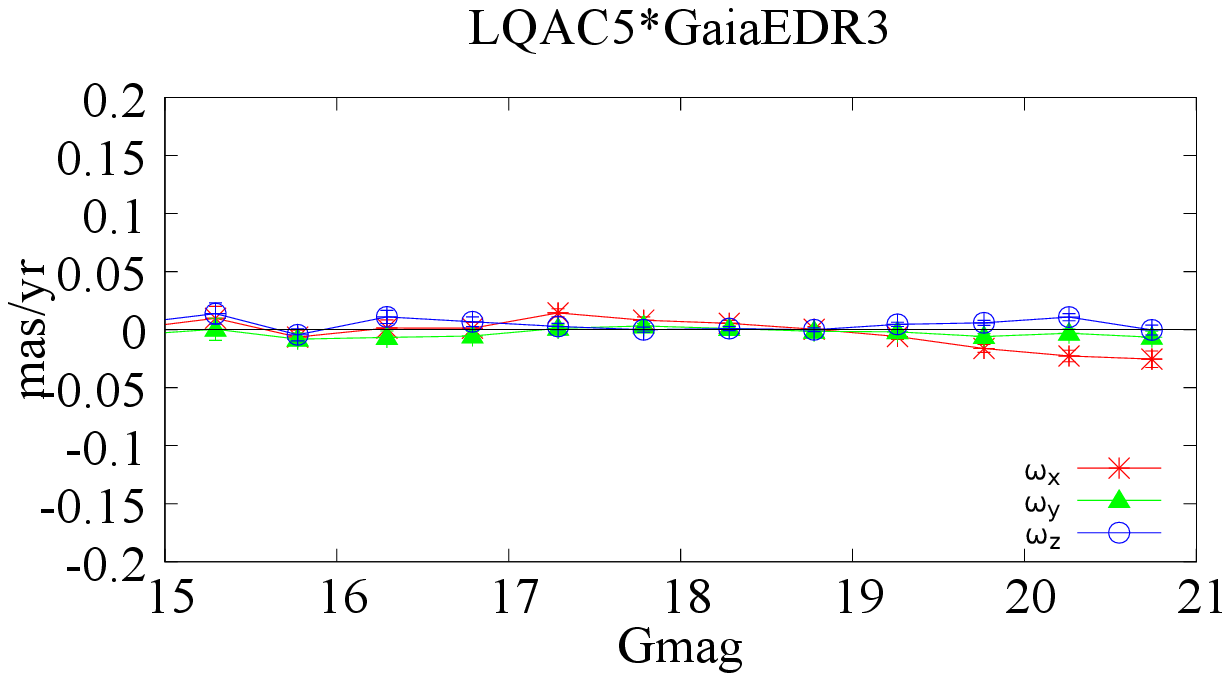}
 \includegraphics[width = 58mm,]{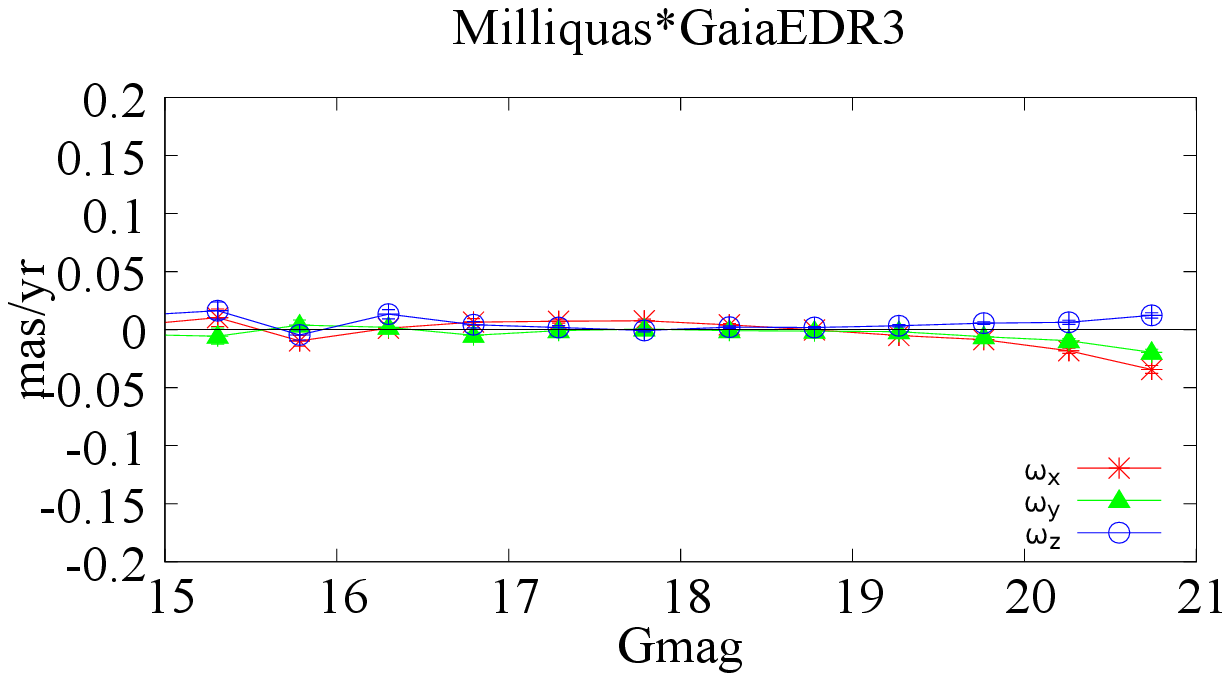}
\caption{Components $\omega_{x}$ (asterisk),  $\omega_{y}$(triangles)  and $\omega_{z}$ (open circles) of the angular velocity vector of the {\it Gaia}~EDR3 system with respect to system implemented by {\it Gaia}~EDR3-positions of extragalactic sources from AllWISEAGN, Milliquas and LQAC5 catalogues as a function of G stellar magnitude from {\it Gaia}~EDR3.}
\label{fig3}
 \includegraphics[width = 58mm,]{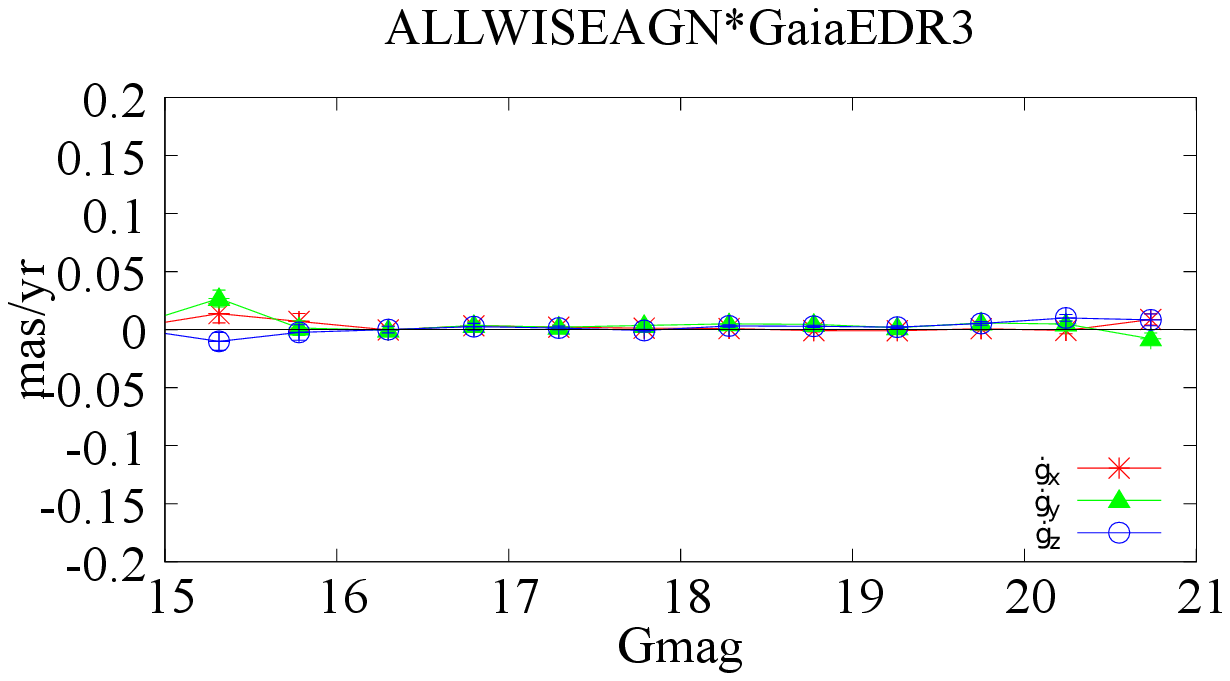}
 \includegraphics[width = 58mm,]{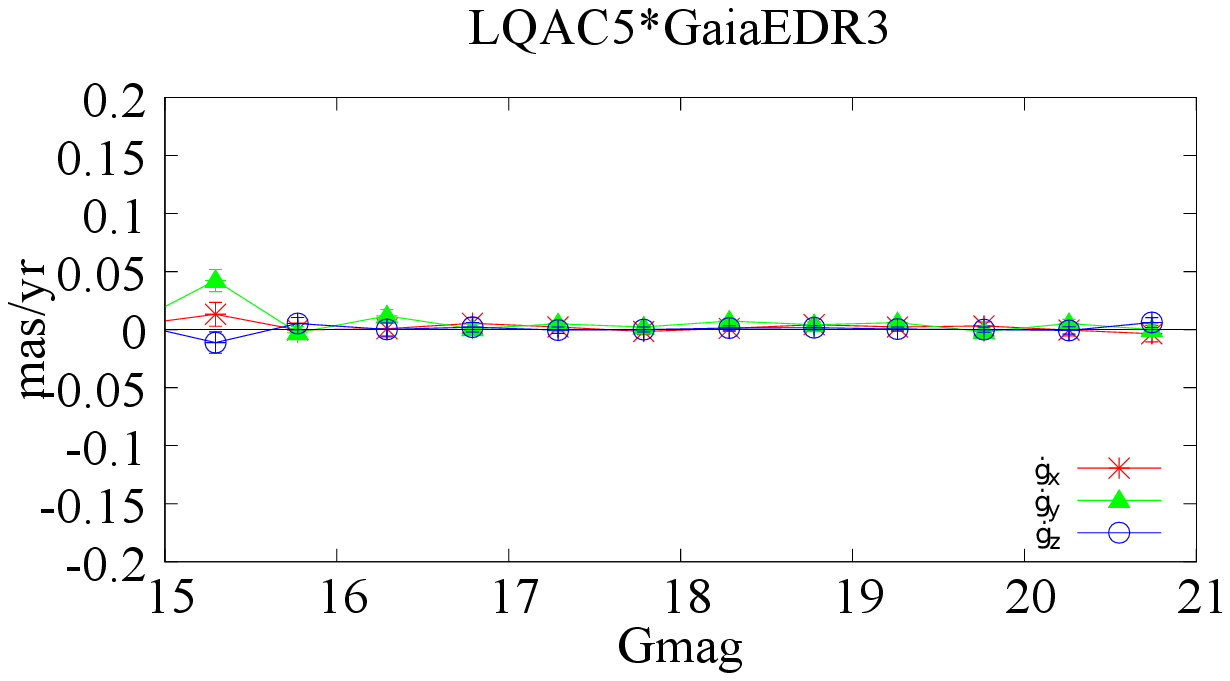}
 \includegraphics[width = 58mm,]{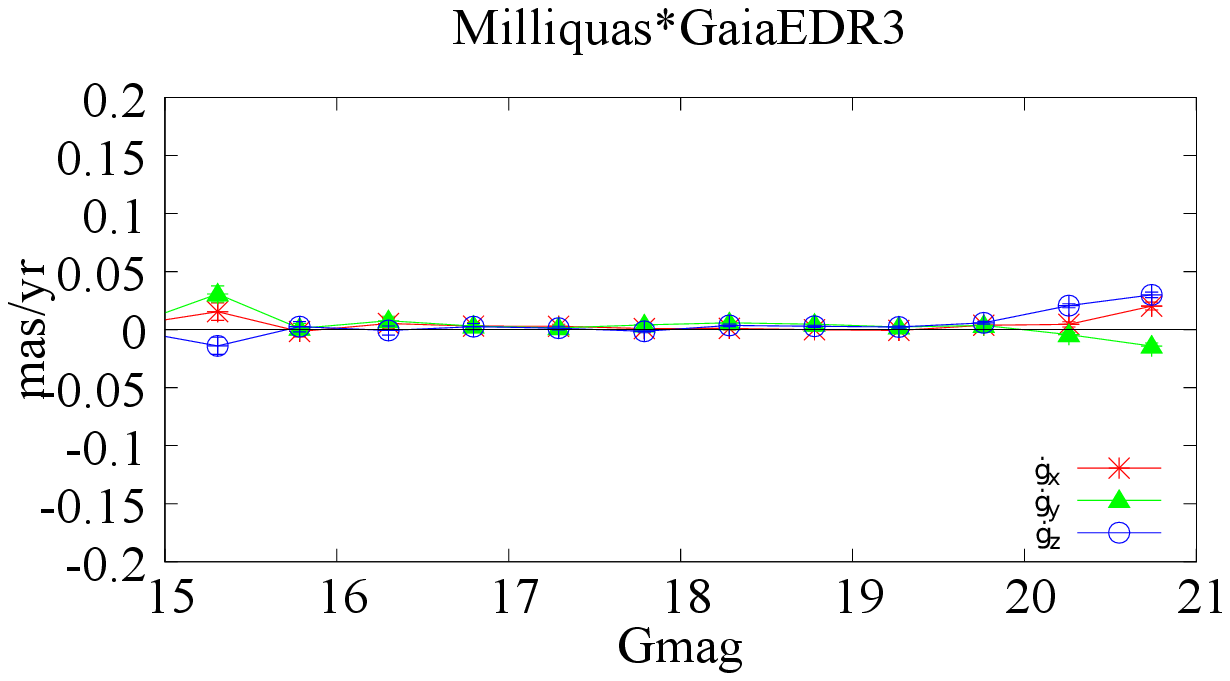}
\caption{Components $\dot{g}_{x}$ (asterisk), $\dot{g}_{y}$ (triangles)  and $\dot{g}_{z}$ (open circles) of the displacement velocity vector of the {\it Gaia}EDR3 system with respect to system implemented by {\it Gaia}~EDR3-positions of extragalactic sources from AllWISEAGN, Milliquas and LQAC5 catalogues as a function of G stellar magnitude from {\it Gaia}EDR3.}
\label{fig4}
\end{figure*}

\begin{figure*}
\includegraphics[width = 58mm,]{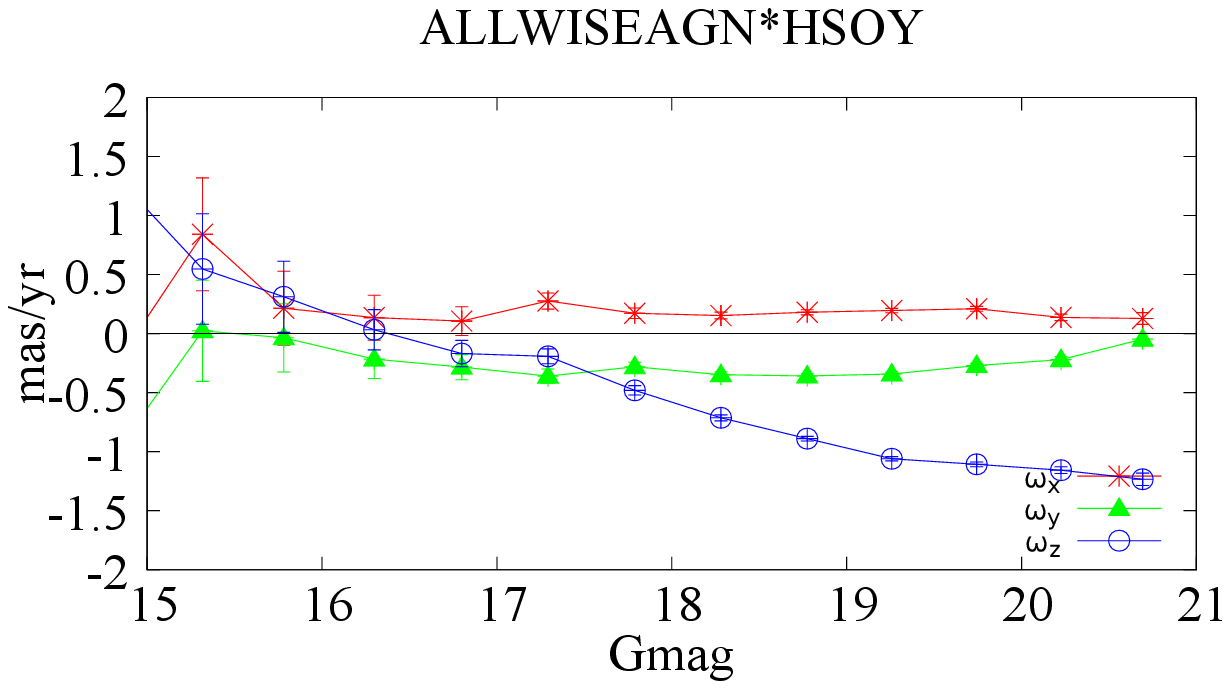}
\includegraphics[width = 58mm,]{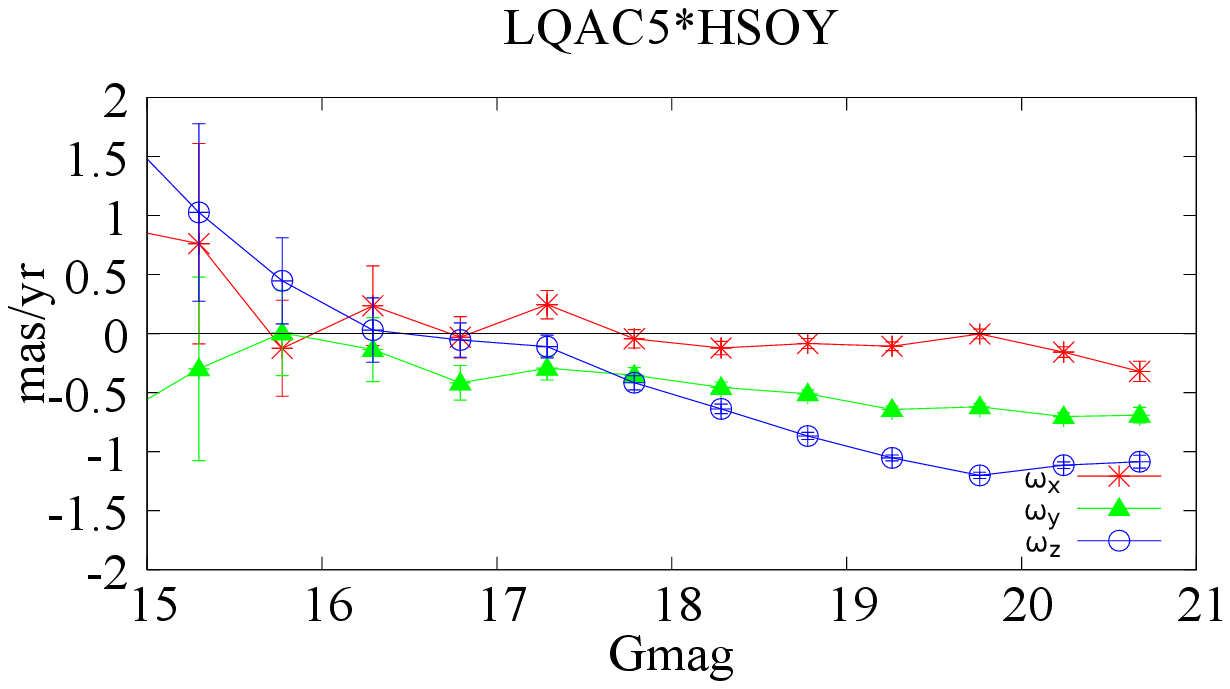}
\includegraphics[width = 58mm,]{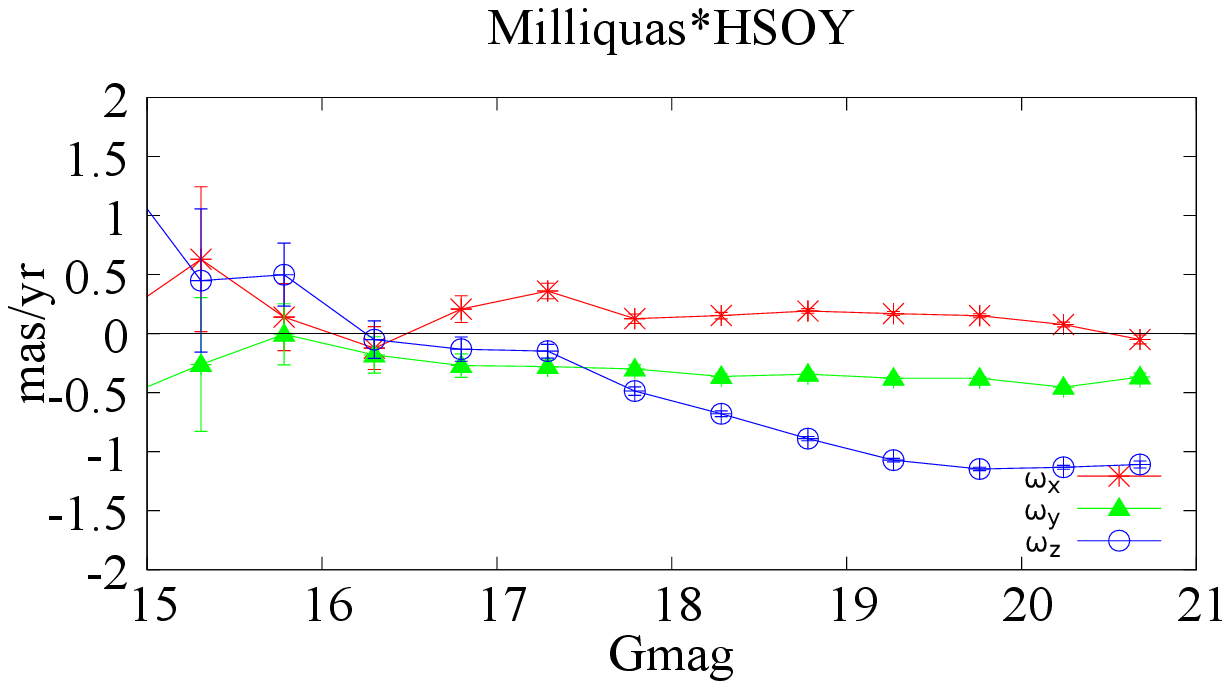}
\caption{Components $\omega_{x}$ (asterisk),  $\omega_{y}$(triangles)  and $\omega_{z}$ (open circles) of the angular velocity vector of the HSOY system with respect to system of implemented by {\it Gaia}~EDR3-positions of extragalactic sources from AllWISEAGN, Milliquas and LQAC5 catalogues as a function of G stellar magnitude from {\it Gaia}EDR3.}
\label{fig5}
\includegraphics[width = 58mm,]{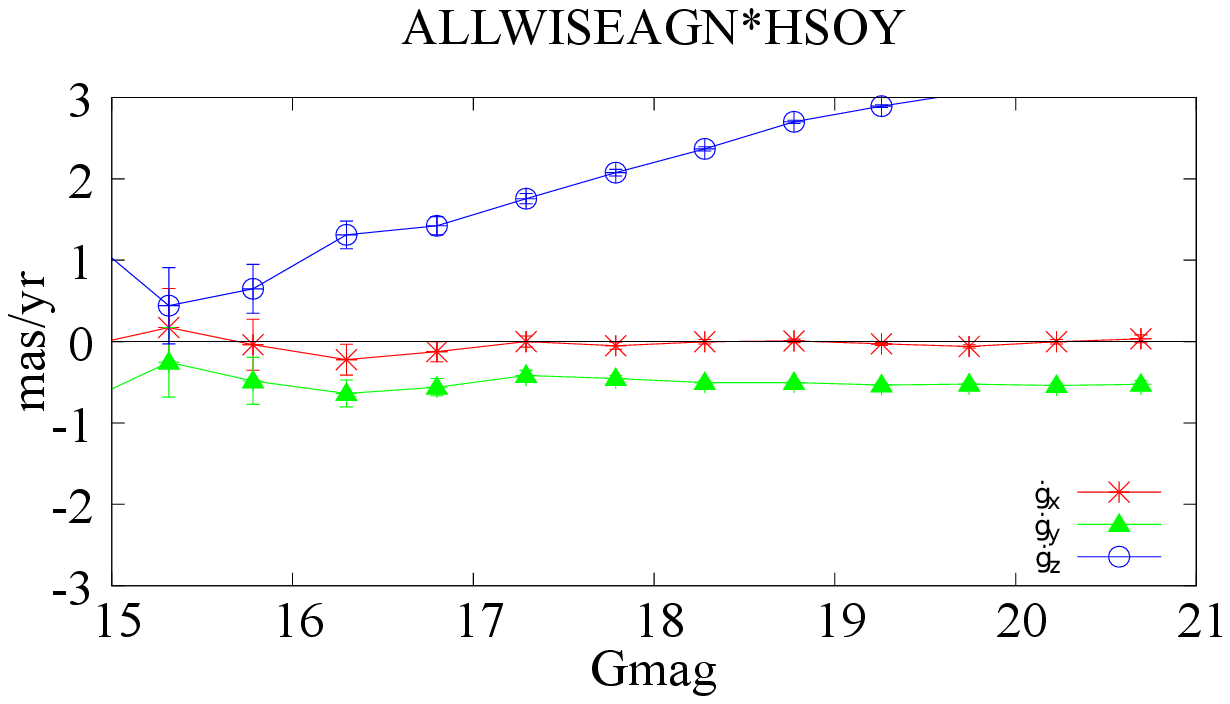}
\includegraphics[width = 58mm,]{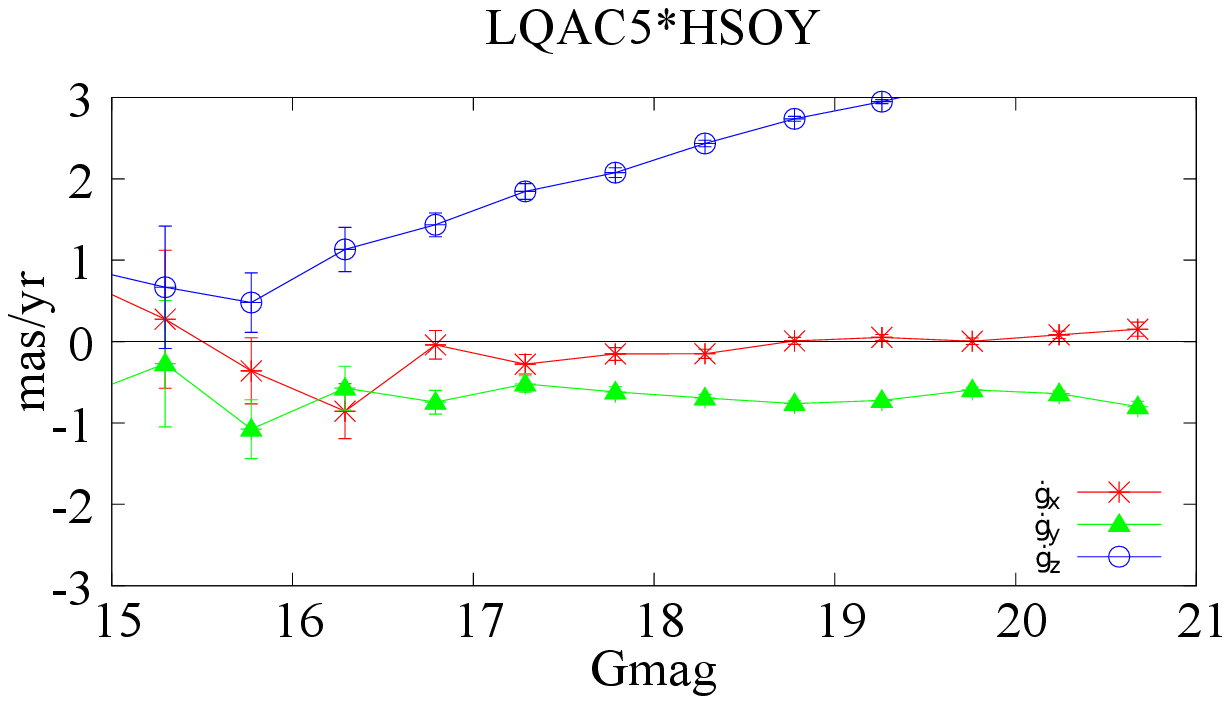}
\includegraphics[width = 58mm,]{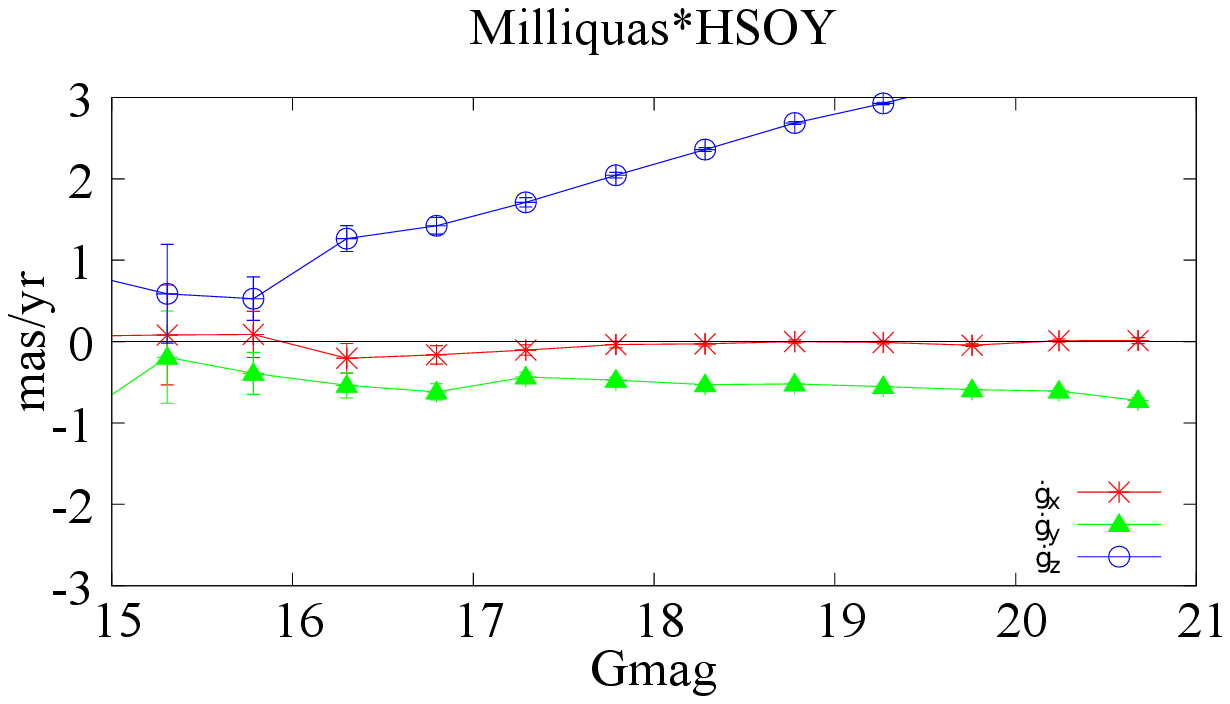}
\caption{Components $\dot{g}_{x}$ (asterisk), $\dot{g}_{y}$ (triangles)  and $\dot{g}_{z}$ (open circles) of the displacement velocity vector of the HSOY system with respect to system of implemented by {\it Gaia}~EDR3-positions of extragalactic sources from AllWISEAGN, Milliquas and LQAC5 catalogues as a function of G stellar magnitude from {\it Gaia}EDR3.}
\label{fig6}
\end{figure*}

\begin{figure*}
\includegraphics[width = 58mm,]{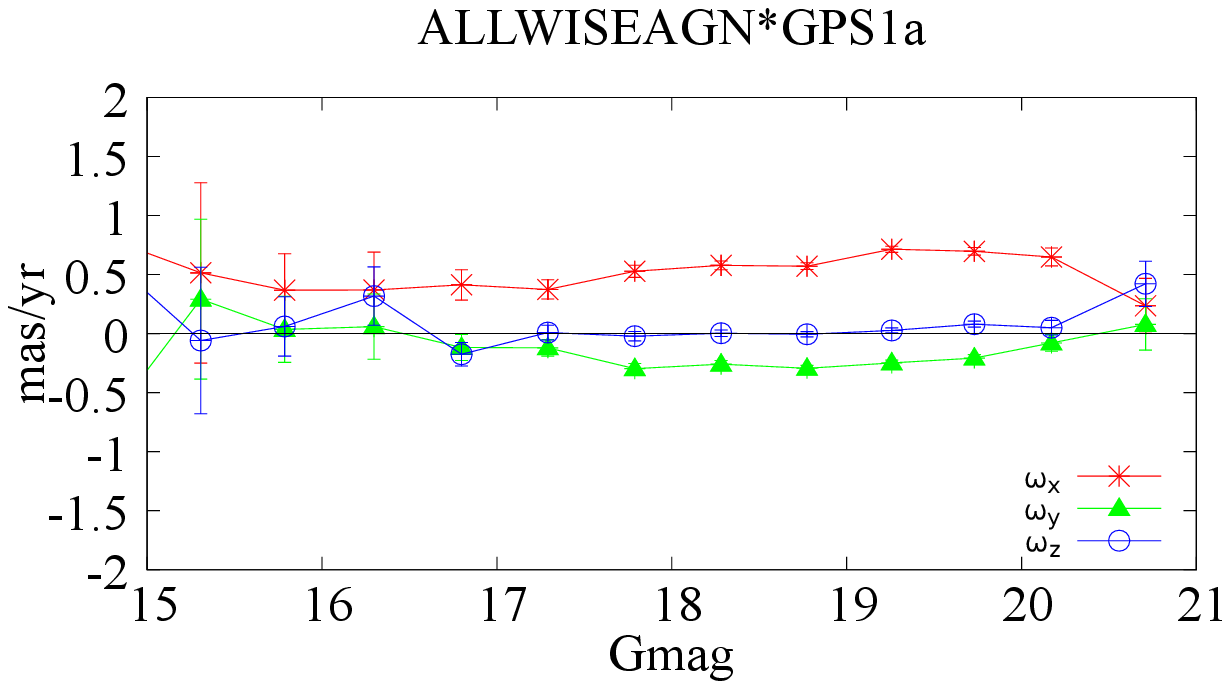}
\includegraphics[width = 58mm,]{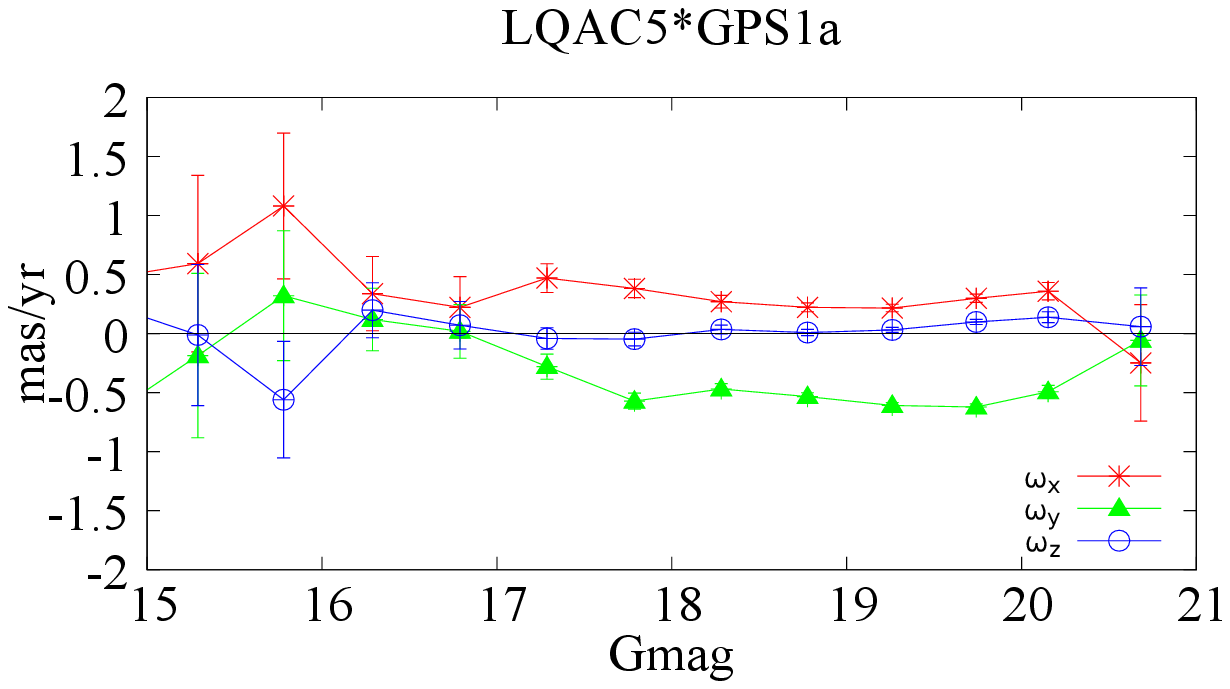}
\includegraphics[width = 58mm,]{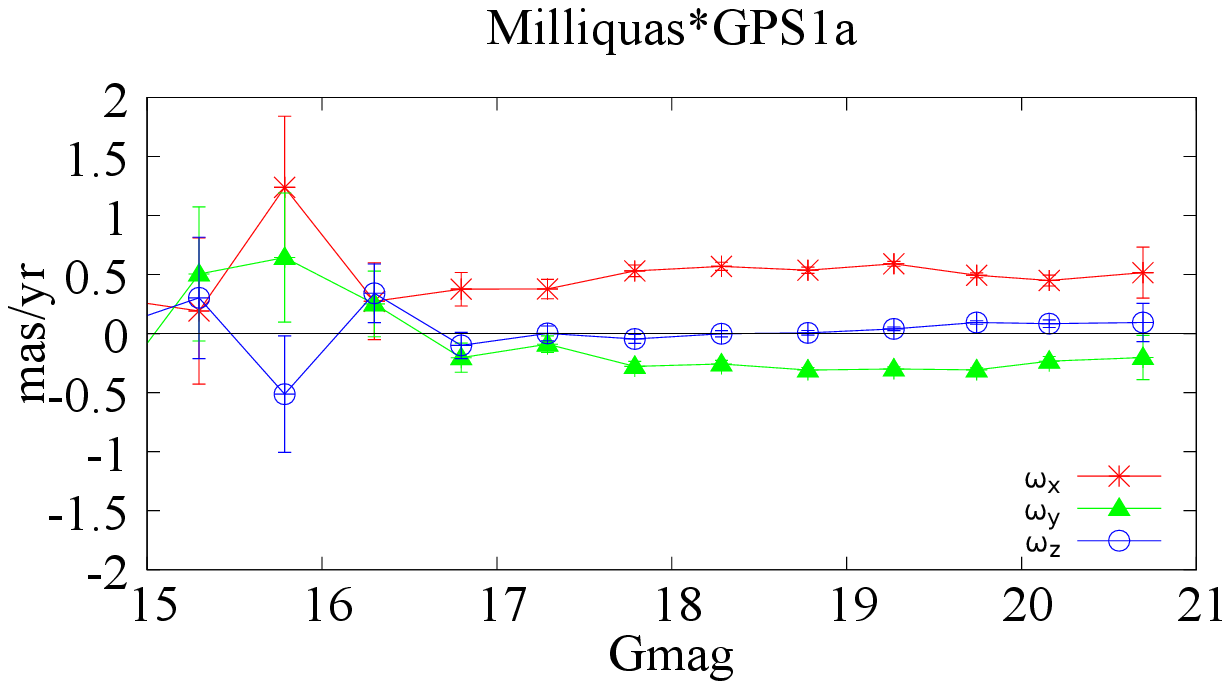}
\includegraphics[width = 58mm,]{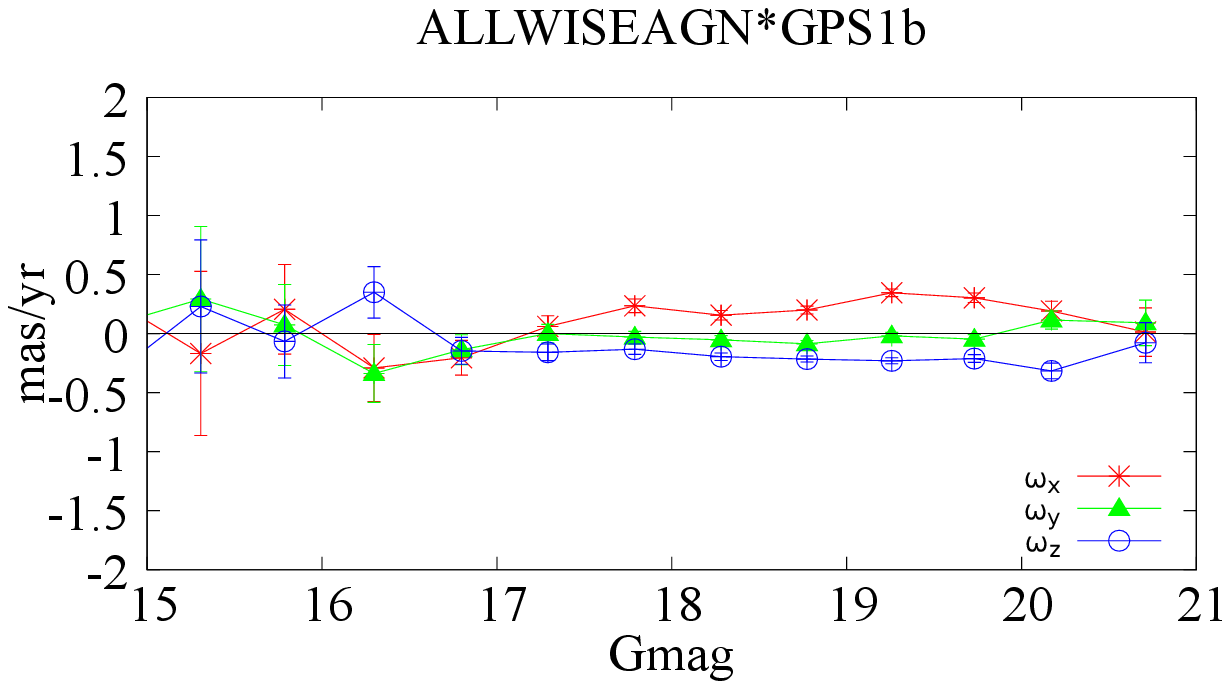}
\includegraphics[width = 58mm,]{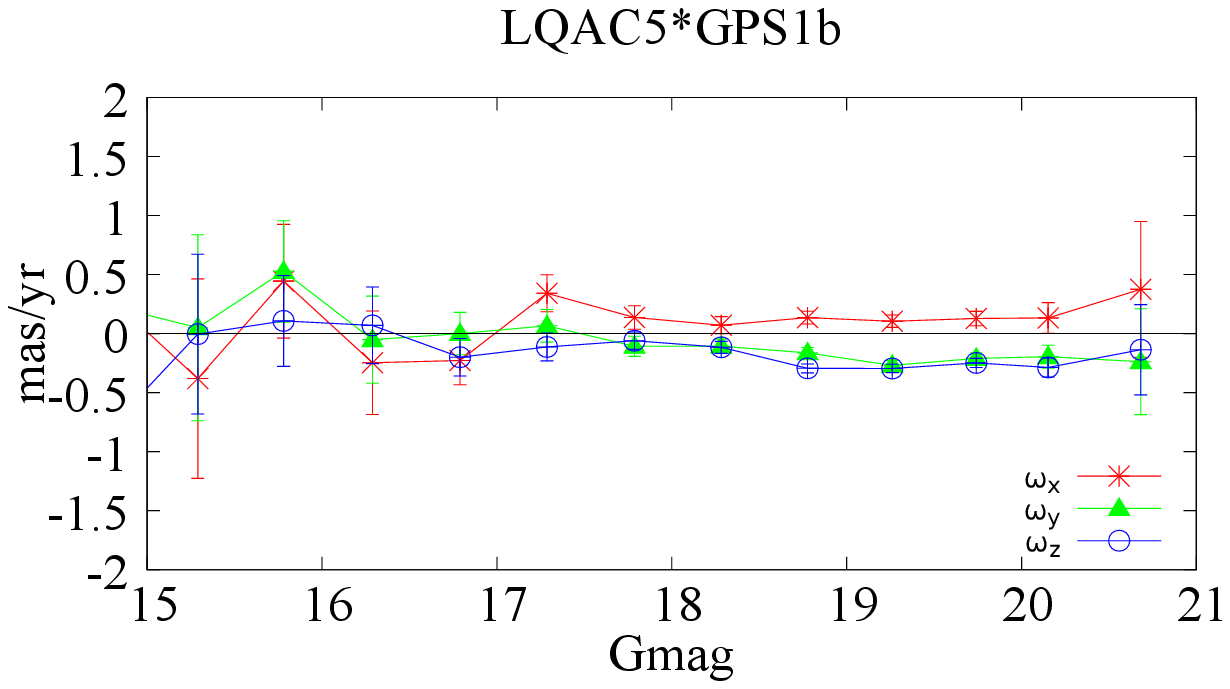}
\includegraphics[width = 58mm,]{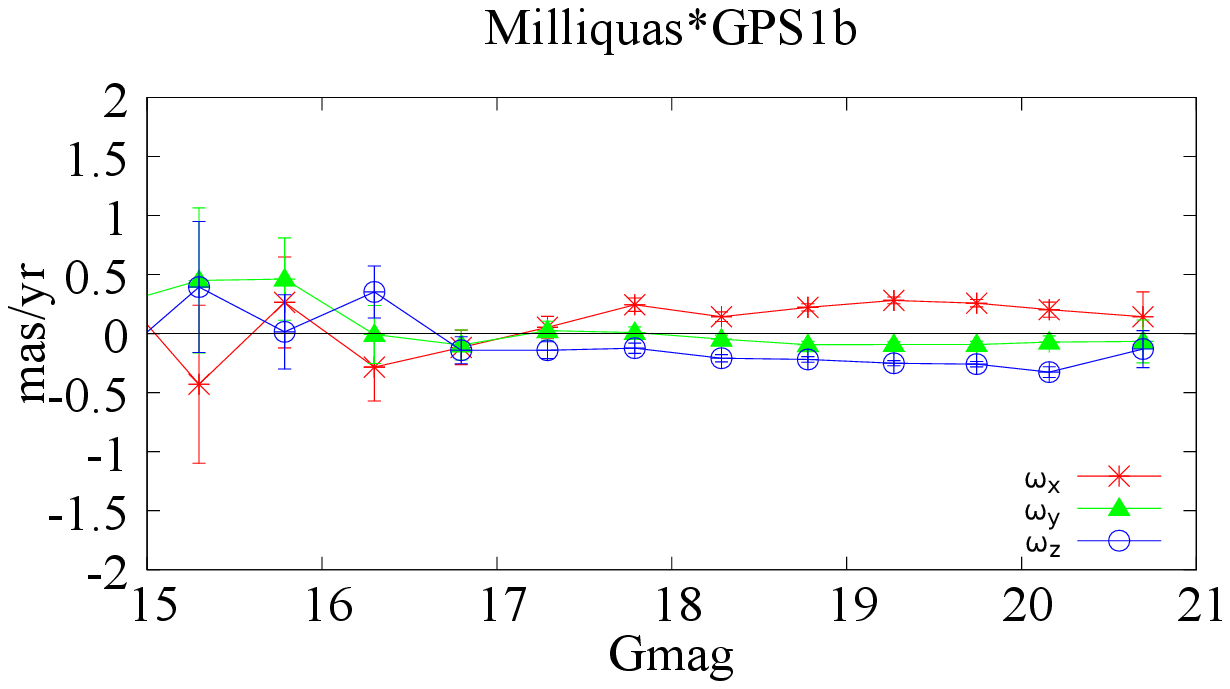}
\caption{Components $\omega_{x}$ (asterisk),  $\omega_{y}$(triangles)  and $\omega_{z}$ (open circles) of the angular velocity vector of the GPS1a and GPS1b system with respect to system of implemented by {\it Gaia}~EDR3-positions of extragalactic sources from AllWISEAGN, Milliquas and LQAC5 catalogues as a function of G stellar magnitude from {\it Gaia}~EDR3.}
\label{fig7}
\includegraphics[width = 58mm,]{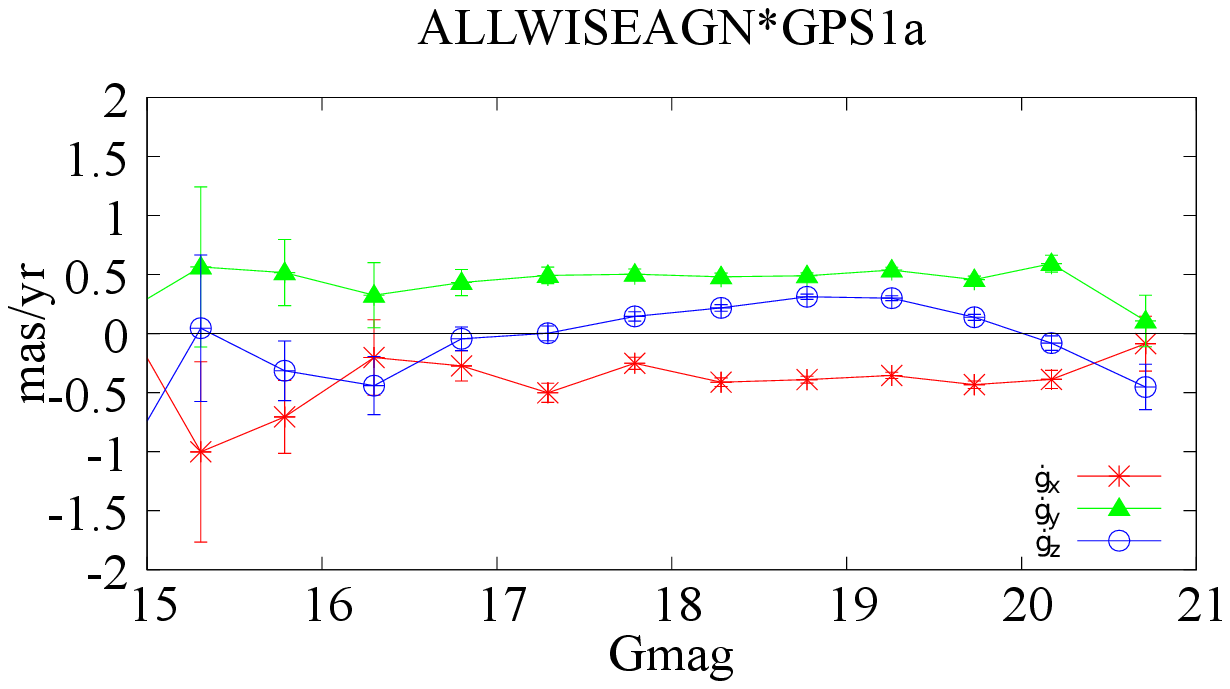}
\includegraphics[width = 58mm,]{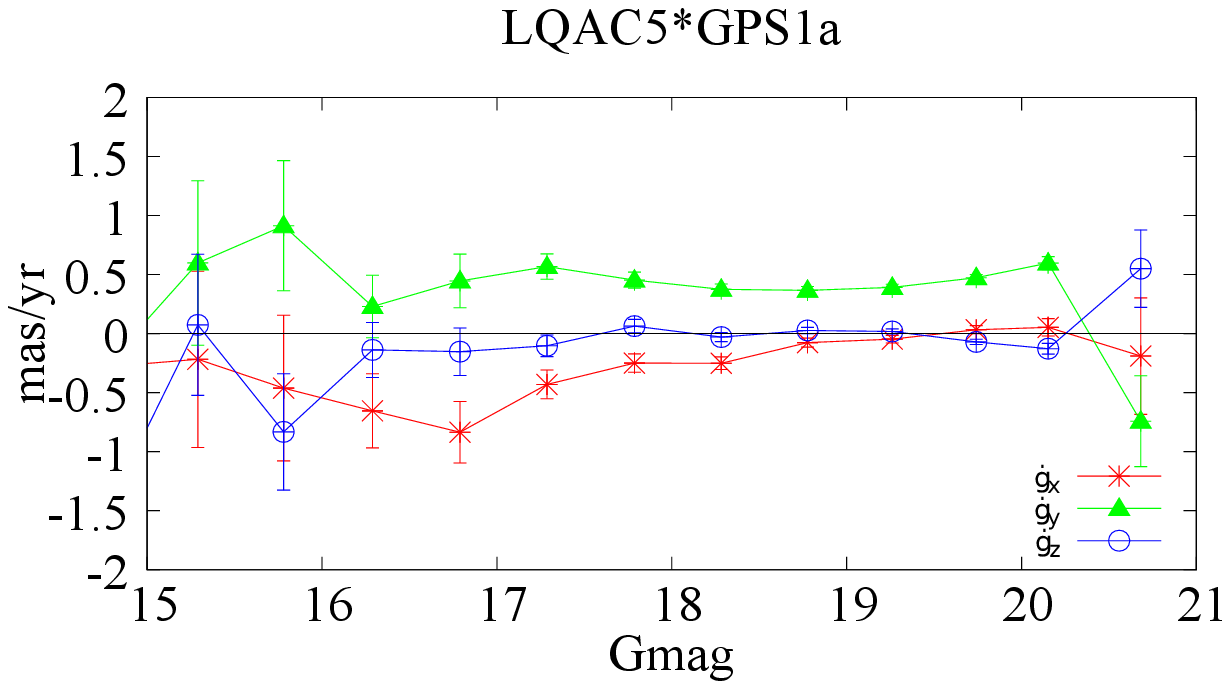}
\includegraphics[width = 58mm,]{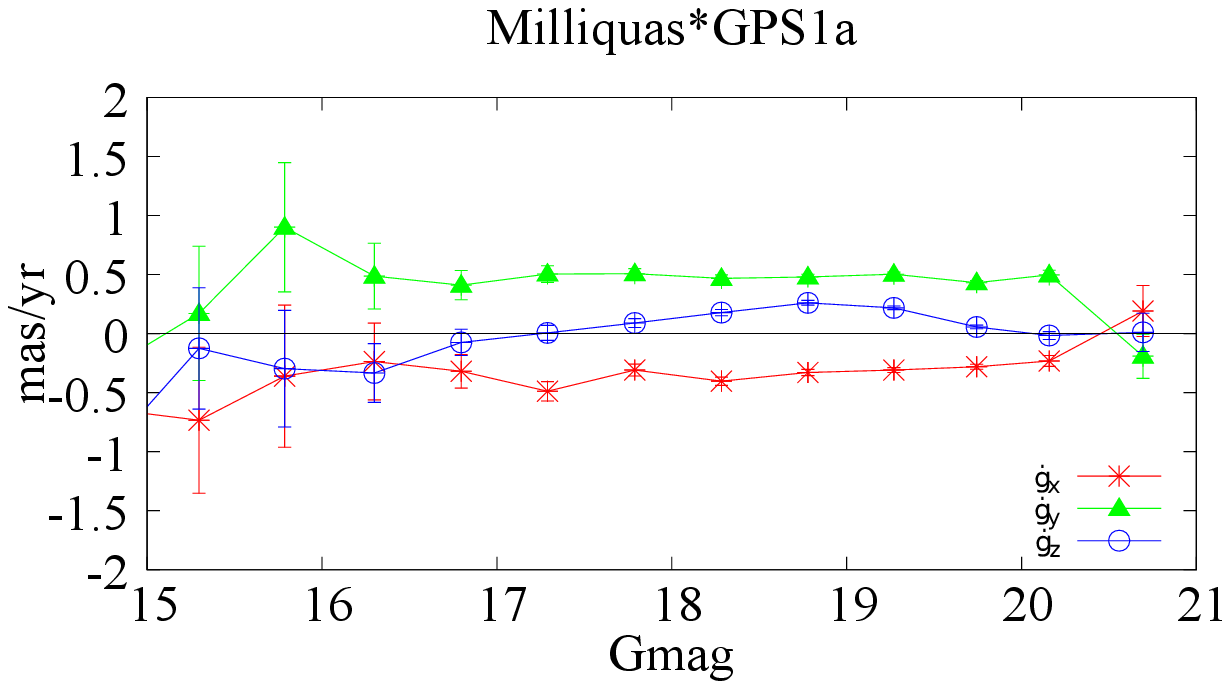}
\includegraphics[width = 58mm,]{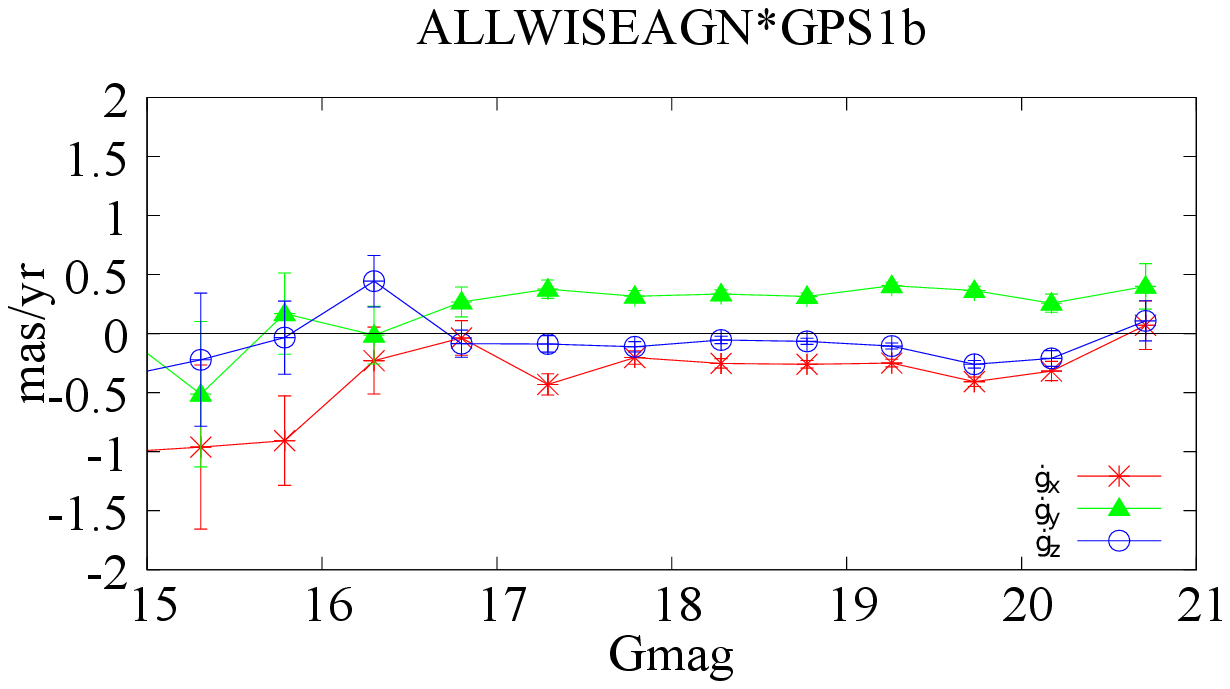}
\includegraphics[width = 58mm,]{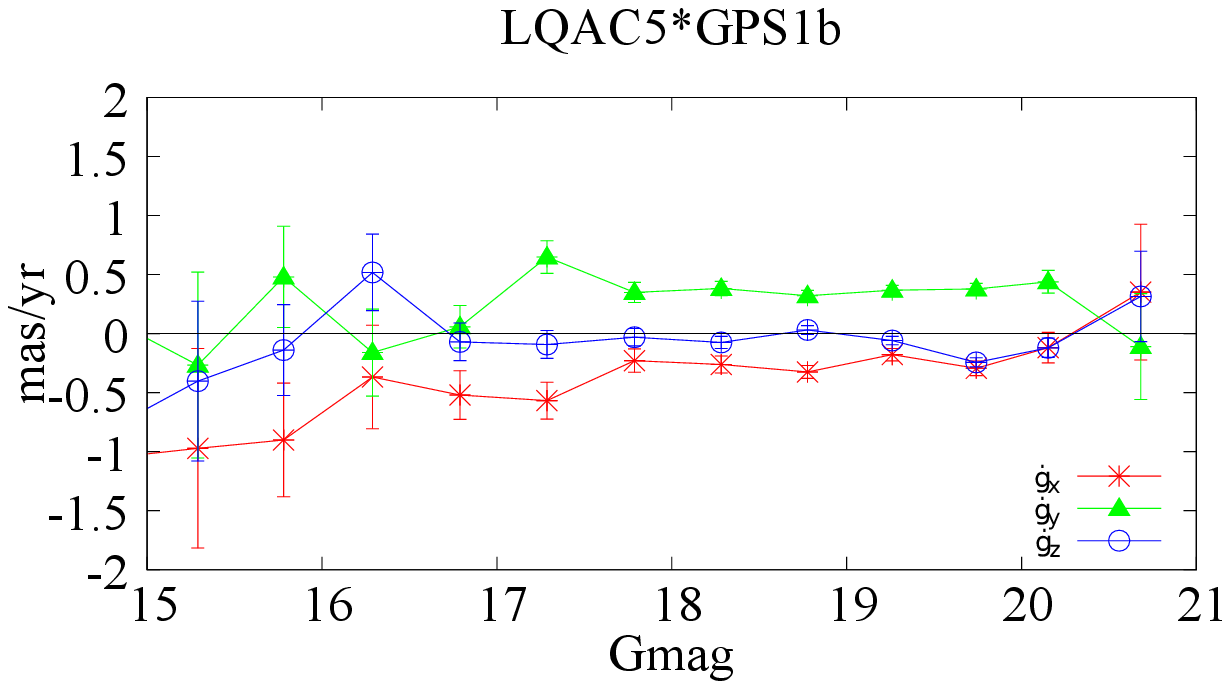}
\includegraphics[width = 58mm,]{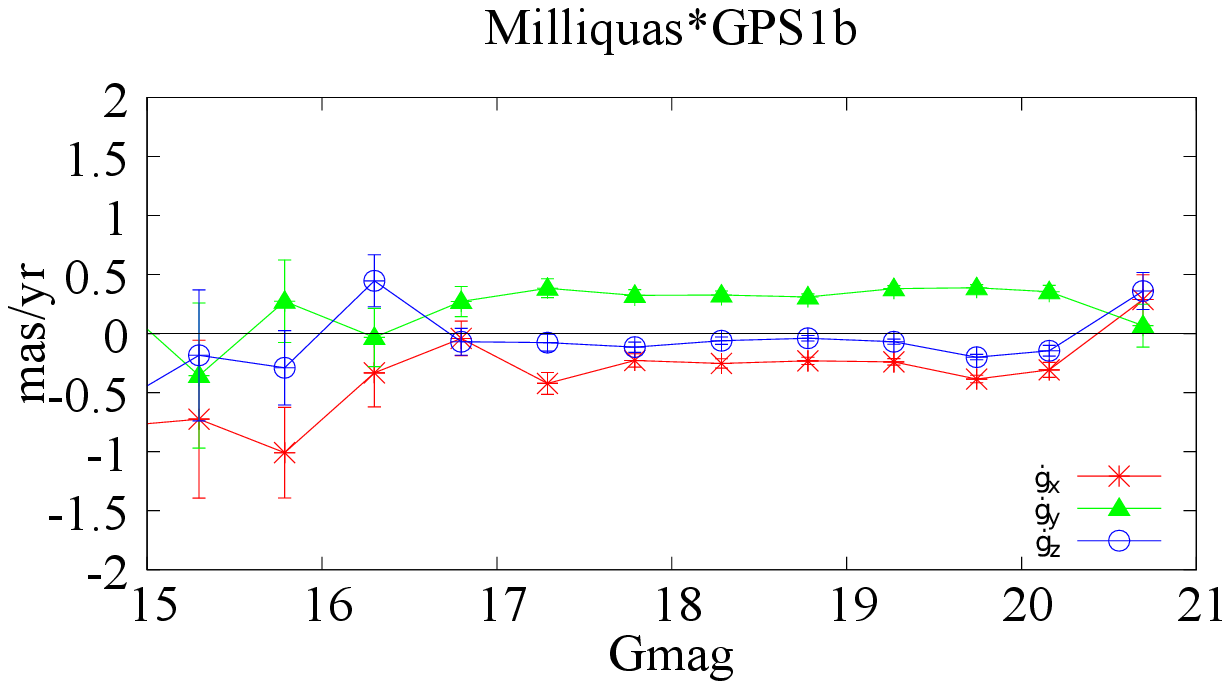}
\caption{Components $\dot{g}_{x}$ (asterisk), $\dot{g}_{y}$ (triangles)  and $\dot{g}_{z}$ (open circles) of the displacement velocity vector of the GPS1a and GPS1b system with respect to system of implemented by {\it Gaia}~EDR3-positions of extragalactic sources from AllWISEAGN, Milliquas and LQAC5 catalogues as a function of G stellar magnitude from {\it Gaia}~EDR3.}
\label{fig8}
\end{figure*}

\begin{figure*}
\vspace*{0pt}
\includegraphics[width = 58mm,]{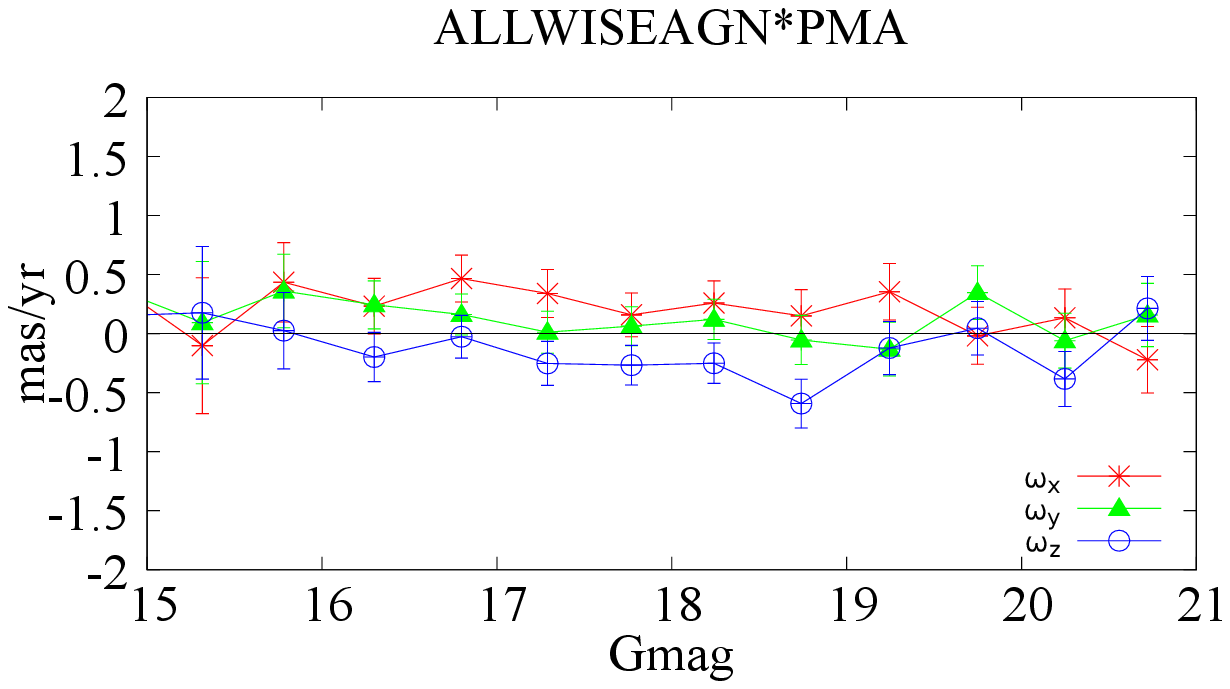}
\includegraphics[width = 58mm,]{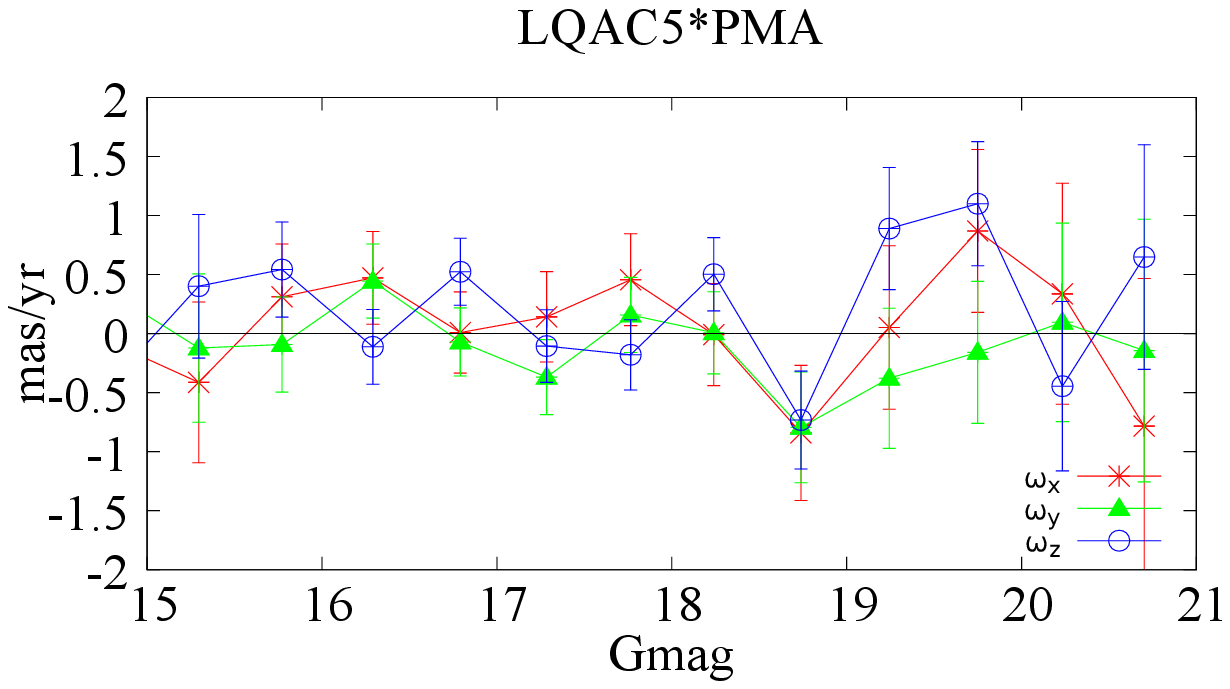}
\includegraphics[width = 58mm,]{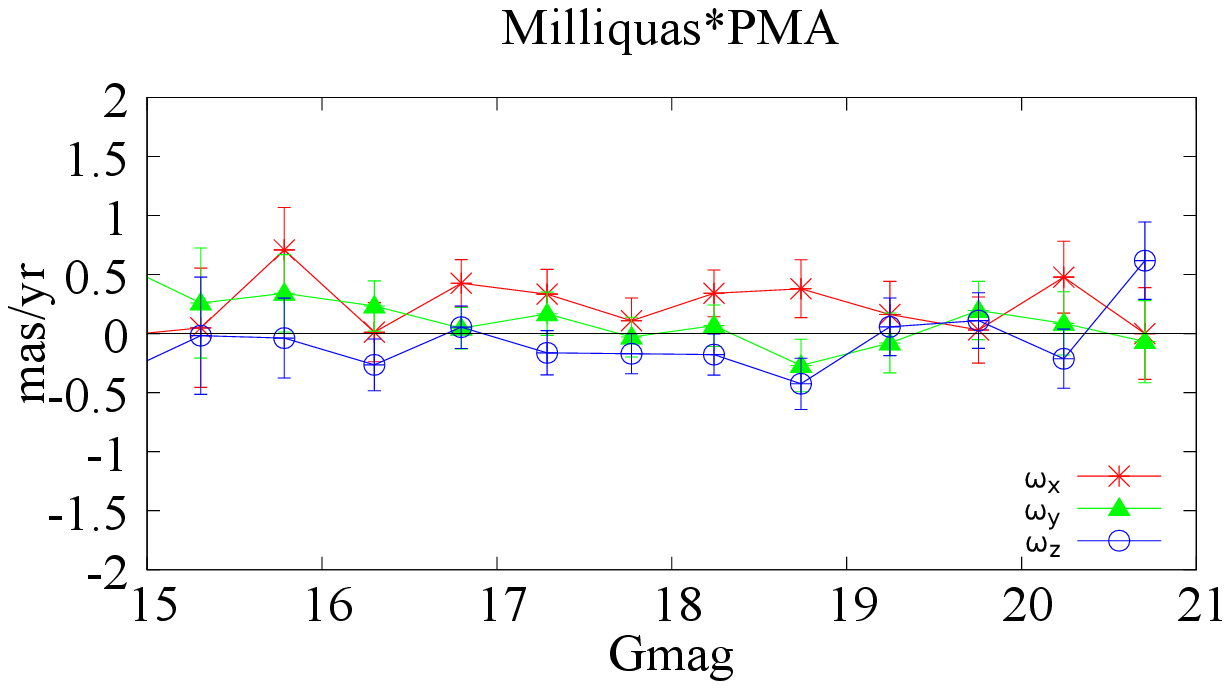}
\caption{Components $\omega_{x}$ (asterisk),  $\omega_{y}$(triangles)  and $\omega_{z}$ (open circles) of the angular velocity vector of the PMA system with respect to system of implemented by {\it Gaia}~EDR3-positions of extragalactic sources from AllWISEAGN, Milliquas and LQAC5 catalogues as a function of G stellar magnitude from {\it Gaia}~EDR3.}
\label{fig9}
\includegraphics[width = 58mm,]{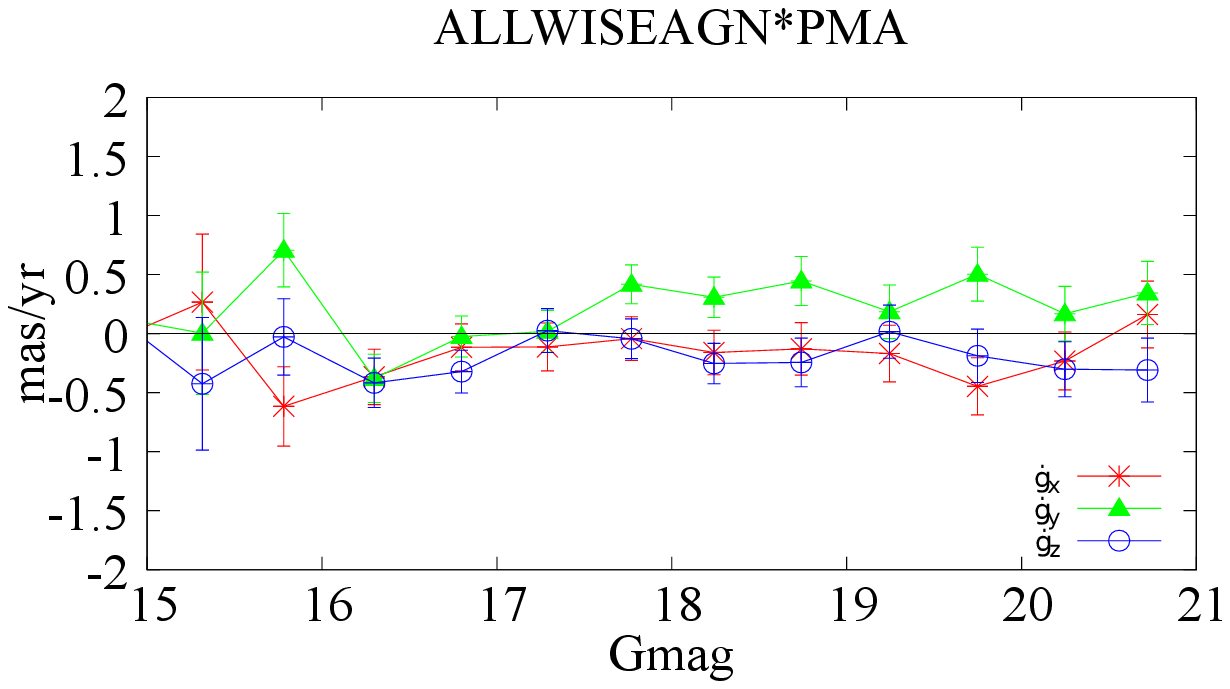}
\includegraphics[width = 58mm,]{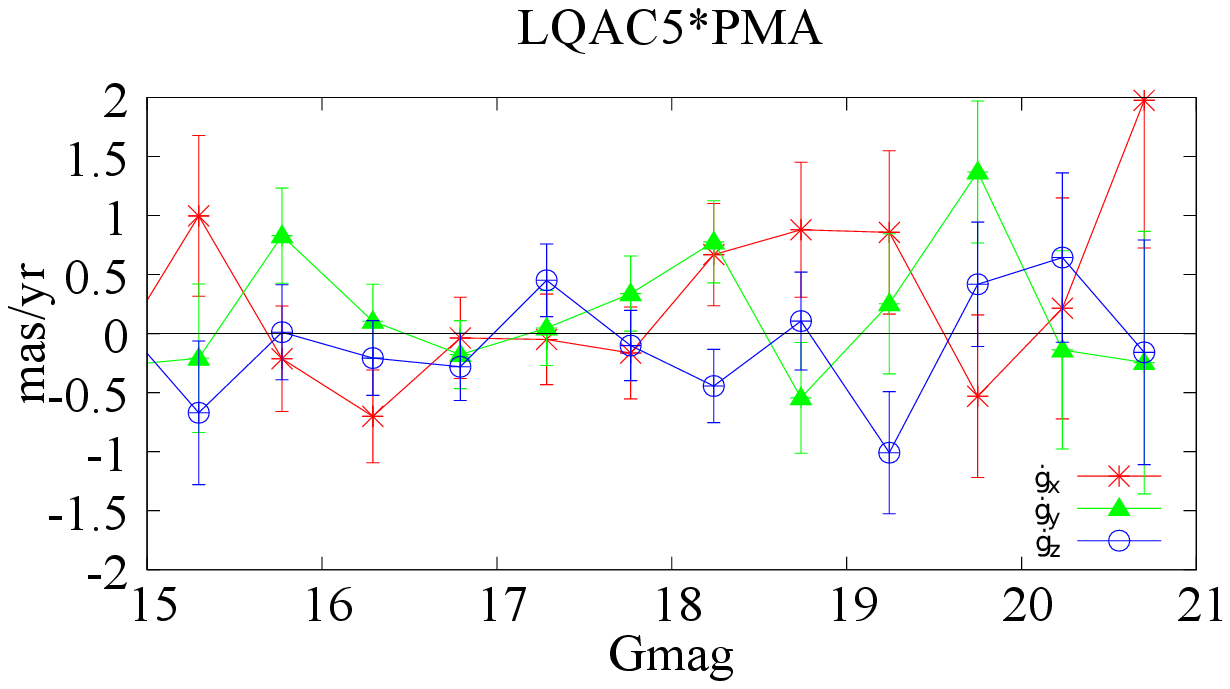}
\includegraphics[width = 58mm,]{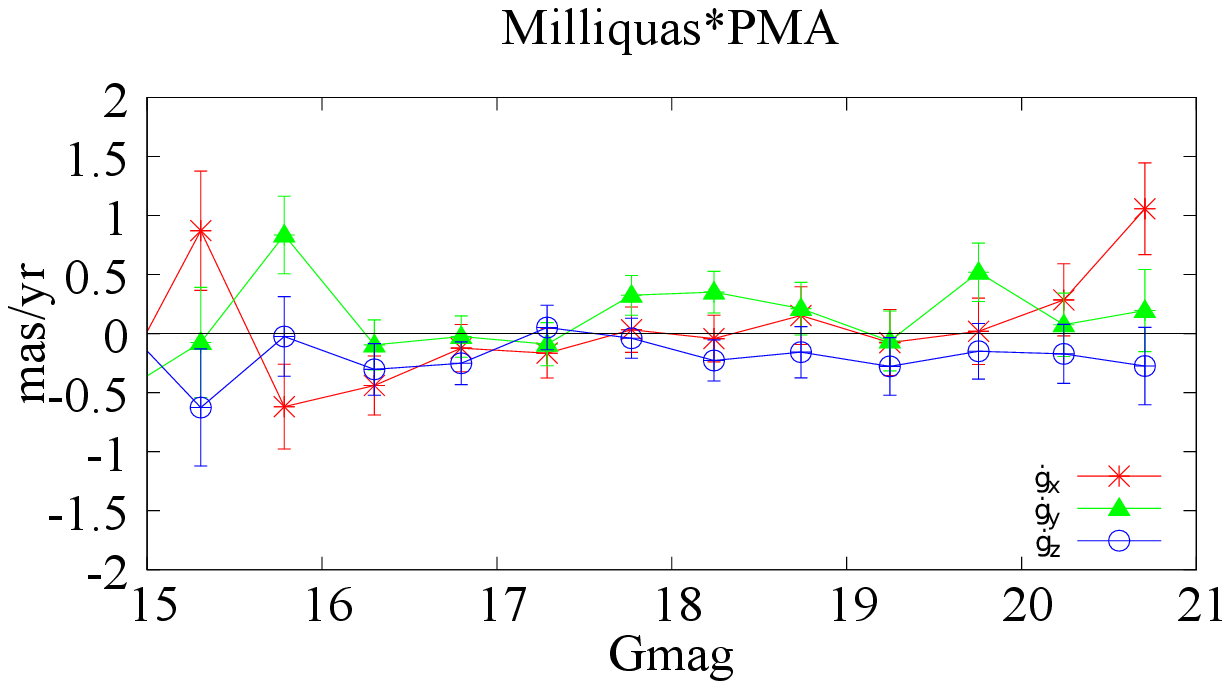}
\caption{Components $\dot{g}_{x}$ (asterisk), $\dot{g}_{y}$ (triangles)  and $\dot{g}_{z}$ (open circles) of the displacement velocity vector of the PMA system with respect to system of implemented by  {\it Gaia}~EDR3-positions of extragalactic sources from AllWISEAGN, Milliquas and LQAC5 catalogues as a function of G stellar magnitude from {\it Gaia}~EDR3.}
\label{fig10}
\vspace*{0.1 cm}
\end{figure*}

The AVV and DVV components were calculated by the least squares method from the formal proper motions of the selected extragalactic sources contained in the {\it Gaia}~EDR3, HSOY, GPS1 and PMA catalogues. Figs. ~\ref{fig3} -~\ref{fig10}~ show dependence the AVV and DVV components on stellar magnitude from {\it Gaia}~EDR3 catalogue. These components of the coordinate systems implemented by the above catalogues, relative to extragalactic sources from AllWISEAGN, Milliquas and LQAC5. It can be seen from these figures that the {\it Gaia}~EDR3 coordinate system has practically no rotation and glide relative to the sources from  AllWISEAGN, Milliquas and LQAC5 in the range from 15 to 21 stellar magnitude.

Further, we consider the rotation and glide of the coordinate systems of the HSOY, PMA and GPS1 catalogues relative to the extragalactic sources from AllWISEAGN, Milliquas and LQAC5, which are contained in these catalogues. The UCAC5 and TGAS catalogues are not analysed due to the almost complete absence of extragalactic sources in them.

The reference frame realised by the positions and formal proper motions of extragalactic sources from the HSOY catalogue shows a pronounced rotation and glide relative to the sources from AllWISEAGN and Milliquas. The components of these vectors are very similar to each other in behaviour, and reach values of $\pm$2~mas~yr$^{-1}$. As compared to the LQAC5 sources, the AVV and DVV components are somewhat lower in amplitude in the range from 15 to 17 stellar G magnitude. It is clearly seen from the Fig.~\ref{fig5} and Fig.~\ref{fig6} that the behaviours of components $\omega_{x}$, $\dot{g}_{x}$ and $\omega_{y}$, $\dot{g}_{y}$ are in better agreement with each other within $\pm$0.5~mas~yr$^{-1}$, while component $\omega_{z}$ and $\dot{g}_{z}$ differs from them both in behaviour and in magnitude, reaching a value of 1.5 -- 3 mas yr$^{-1}$.

Considering the AVV and DVV components of the coordinate systems of the GPS1a and GPS1b catalogues with respect to the extragalactic sources contained in AllWISEAGN, Milliquas and LQAC5, the GPS1b coordinate system can be seen to have slightly smaller values of the components than those of GPS1a. The values of the AVV and DVV components of both GPS1a and GPS1b coordinate systems, obtained from the formal proper motions of extragalactic objects, are much less than the analogous values characterising the rotation and glide of the GPS1a and GPS1b coordinate systems relative to the {\it Gaia}~EDR3 stellar frame (Fig. ~\ref{fig1} and Fig.~\ref{fig2}). This means that the stellar reference frames of the GPS1a and GPS1b catalogues, realised by the positions and proper motions of their stars, are not consistent with their extragalactic reference frames, realised by the directions to the AllWISEAGN, Milliquas, and LQAC5 sources. As a result, zero-points of the systems of proper motions of stars in the GPS1a and GPS1b catalogues turn out to be shifted relative to the extragalactic reference frames. This means that the systems of proper motions of stars in these catalogues contain systematic errors that cannot be eliminated even if they are eliminated in the coordinate system implemented  by extragalactic sources.

Relative to the extragalactic sources of AllWISEAGN and Milliquas, the AVV and DVV components of the PMA coordinate system in the range from 16.5 to 19.5 stellar G magnitude practically do not depend on the magnitude Fig. ~\ref{fig9}, Fig.~\ref{fig10}  and do not exceed the value $\pm$ 0.3~mas~yr$^{-1}$ as declared in \citep{a1}. Note that the procedure for absolutizing the proper motions of stars of the PMA catalogue was carried out using the positions of 1.6 million extragalactic sources without splitting them into quasars and galaxies. The proper motions of all objects in the PMA catalogue, including extragalactic sources (with the exception of those used for the absolutization procedure), were determined with the same procedure.

The values of the AVV and DVV components relative to the LQAC5 quasars turned out to be somewhat different from the values  indicated above, reaching values of about 1 mas yr$^{-1}$ at the 19 stellar magnitude. We believe that it is related with small number of quasars contained in the PMA catalogue. Note that the proper motions of the PMA objects were derived from a combination of the positions of the {\it Gaia}~DR1 survey and 2MASS catalogue -- near-infrared survey that contains a small number of blue objects in the range of faint magnitudes, - only $\sim$ 28,000 quasars. The bar sizes shown in Fig.~\ref{fig9} and Fig.~\ref{fig10} confirm indirectly the small number of quasars used for averaging.

We also note that the AVV and DVV components obtained from stars and extragalactic sources do not reveal any dependence on the magnitude and virtually do not differ at an accuracy level of 0.5 mas yr$^{-1}$. This result indicates the consistency of the system of proper motions of stars and the system of formal proper motions of extragalactic sources.

\section{Conclusions}

There are several modern catalogues of proper motions of stars obtained by a combination of ground-based observations and data from the {\it Gaia} space mission. Comparison of the proper motions of the stars of these catalogues with the data of the {\it Gaia} Early Data Release 3 made it possible to determine the components of mutual rotation and displacement of the coordinate systems specified by these catalogues. It turned out that only the values of the AVV and DVV components between the coordinate systems indicated by the PMA and {\it Gaia}~EDR3 data do not exceed 0.2 mas yr$^{-1}$ in the range from 10 to 20 stellar magnitude. As for the values of similar components obtained from a comparison of the GPS1 and HSOY data with the data of the {\it Gaia}~EDR3 catalogue, they are in the range of 0.5 -- 3.5 mas yr$^{-1}$ and reveal dependence on magnitude. These facts indicate that the coordinate systems implemented by the GPS1 and HSOY catalogues have residual fictitious rotation, and their systems of proper motions of stars are burdened with systematic errors.

We have performed the analysis of the formal proper motions of extragalactic objects from ALLWISEAGN, Milliquas and LQAC5 contained in the {\it Gaia}~DR3, HSOY, GPS1 and PMA catalogues, has also shown that only the coordinate systems defined by the {\it Gaia}~DR3 and PMA data in the range from 16 to 20 stellar magnitude have no rotation and glide relative to extragalactic objects within $\pm$ 0.15 and $\pm$ 0.3~mas~yr$^{-1}$ respectively.

\section{Acknowledgements}
This work has made use of data from the European Space Agency (ESA) mission {\it Gaia} (\url {https://www.cosmos.esa.int/gaia}), processed by the {\it Gaia} Data Processing and Analysis Consortium (DPAC,\url {https://www.cosmos.esa.int/web/gaia/dpac/consortium}). Funding for the DPAC has been provided by national institutions, in particular the institutions participating in the {\it Gaia} Multilateral Agreement.
This publication makes use of data products from the Wide-field Infrared Survey Explorer, which is a joint project of the University of California, Los Angeles, and the Jet Propulsion Laboratory/California Institute of Technology, and NEOWISE, which is a project of the Jet Propulsion Laboratory/California Institute of Technology. WISE and NEOWISE are funded by the National Aeronautics and Space Administration.
The Pan-STARRS1 Surveys (PS1) and the PS1 public science archive have been made possible through contributions by the Institute for Astronomy, the University of Hawaii, the Pan-STARRS Project Office, the Max-Planck Society and its participating institutes, the Max Planck Institute for Astronomy, Heidelberg and the Max Planck Institute for Extraterrestrial Physics, Garching, The Johns Hopkins University, Durham University, the University of Edinburgh, the Queen’s University Belfast, the Harvard-Smithsonian Center for Astrophysics, the Las Cumbres Observatory Global Telescope Network Incorporated, the National Central University of Taiwan, the Space Telescope Science Institute, the National Aeronautics and Space Administration under Grant No. NNX08AR22G issued through the Planetary Science Division of the NASA Science Mission Directorate, the National Science Foundation Grant No. AST-1238877, the University of Maryland, Eotvos Lorand University (ELTE), the Los Alamos National Laboratory, and the Gordon and Betty Moore Foundation. 

The work of V. Akhmetov and E. Bannikova was supported under the special program of the NRF of Ukraine ”Leading and Young Scientists Research Support” – ”Astrophysical Relativistic Galactic Objects (ARGO): life cycle of active nucleus”, No. 2020.02/0346.

\section*{Data availability}
\addcontentsline{toc}{section}{Data availability}
The used catalogues data are available in a standardised format for readers via the CDS (https://cds.u-strasbg.fr).
The data of cross-matches between catalogues as well as software code used in this paper can be made available upon request by emailing the corresponding author.

\end{document}